\documentclass[useAMS,usenatbib]{mn2e}
\usepackage{deluxetable}
\usepackage{epsf}

\newcommand{\kmsmpc}{\kms\;{\rm Mpc}^{-1}}
\newcommand{\lya}{Ly$\alpha$}
\newcommand{\hkpc}{h^{-1}{\rm kpc}}
\newcommand{\hmpc}{h^{-1}{\rm Mpc}}
\newcommand{\kms}{\;{\rm km}\,{\rm s}^{-1}}
\newcommand{\ion}[2]{{#1~{\sc #2}}}
\newcommand{\gad}{{\sc Gadget-2}}

\title{The Physical Properties and Detectability of Reionization-Epoch Galaxies}

\begin{document}

\author[Dav\'e, Finlator, Oppenheimer]{Romeel Dav\'e$^1$, Kristian Finlator$^1$, Benjamin D. Oppenheimer$^1$\\
$^1$University of Arizona, Steward Observatory, Tucson, AZ 85721
}

\maketitle

 \begin{abstract}
We present predictions drawn from cosmological hydrodynamic simulations
for the physical, photometric and emission line properties of galaxies
present during the latter stages of reionization from $z=9\rightarrow
6$.  We find significant numbers of galaxies that have stellar
masses exceeding $10^9 M_\odot$ during this epoch, with metallicities
exceeding one-thirtieth solar.  Far from primeval ``first-star" objects,
these objects exhibit a significant Balmer break, are likely to have
reionized their infall regions prior to $z=9$, are dominated by atomic
rather than molecular cooling, and are expected to be forming few if any
metal-free stars.  By $z=6$, the space density of $M_*>10^{10}M_\odot$
objects is roughly equivalent to that of luminous red galaxies today.
Galaxies exhibit a slowly evolving comoving autocorrelation length from
$z=9\rightarrow 6$, continuing a trend seen at lower redshifts in which
the rapidly dropping bias counteracts rapidly increasing matter
clustering.  These sources can be marginally detected using current
instruments, but modest increases in sensitivity or survey area would
yield significantly increased samples.  We compare to current observations
of the $z\approx 6$ rest-UV and \lya\ line luminosity functions,
and find good agreement.  We also compare with the $z\sim 7$ object
observed by Egami et al., and find that such systems are ubiquitous in
our simulations.  The intrinsic \lya\ luminosity function evolves slowly
from $z=9\rightarrow 6$, implying that it should also be possible to detect
these objects with upcoming narrow band surveys such as DAzLE, if as we
argue the detectability of \lya\ does not drop significantly to higher
redshifts.  We make predictions for near-IR surveys with {\it JWST},
and show that while a high density of sources will be found, Population
III objects may remain elusive.  We present and compare simulations with
several recipes for superwind feedback, and show that while our broad
conclusions are insensitive to this choice, a feedback model based on
momentum-driven winds is mildly favored in comparisons with available data.
\end{abstract}

 
\section{Introduction}

The epoch of reionization is an important event in the history of our
universe, during which the light of the first stars and galaxies ate
away at the neutral hydrogen fog and transformed the intergalactic medium
(IGM) into the highly ionized cosmic web that we see today~\citep{loe01}.
Detecting this epoch has long been a goal of observational cosmology.
Recent observations of a Gunn-Peterson trough in $z\ga 6$ quasar
spectra suggest that the end of reionization epoch has now been seen
\citep{fan02}.  Conversely, third-year data from the {\it Wilkinson
Microwave Anisotropy Probe} \citep[WMAP;][]{pag06} suggests that the
universe was predominantly ionized out to $z\sim 10$.  When exactly
reionization begins, how it proceeds, and what the sources of the ionizing
photons are remain open questions.

In light of this, a key goal of next generation ground and space-based
telescopes is to detect the objects responsible for reionizing the
universe at $z>6$.  At present these are believed to be star-forming
galaxies since there are insufficient quasars at $z\ga 6$ to do the job
\citep{fan01}, although the latest SDSS samples show a faint-end slope
that is tantalizingly close to the required steepness \citep{fan04}.
Few reionization-epoch objects have been observed up till now
\citep{rho04,ega05,ste05,hu05,cha05}, and hence their detailed physical
properties remain poorly constrained.  In order to optimize detection
strategies with current and future facilities, it is important to
understand the nature of these early galaxies.

Simulations of galaxy formation can provide insights into the expected
properties of these systems within the context of a now well-established
framework for hierarchical structure formation.  While the properties
of dark matter halo assembly are relatively well understood owing to
precisely determined cosmological parameters \citep{spe03,spe06},
the properties of the gas clouds that condense to form observable
stars remain subject to a large range of theoretical uncertainties.
Previous studies have utilized analytic or semi-analytic approaches
to predict the properties of early galaxies, based on simplified
models for the relevant baryonic physics \citep[e.g.][and
references therein]{bark01}.  An alternative approach is to employ
full cosmological hydrodynamic simulations that account for all the relevant
cooling and star formation processes associated with galaxy formation,
as presently understood, constrained by rapidly-improving observations
of high-redshift systems.  Such simulations carry information about the
full three-dimensional structure of accretion and feedback processes
that are vital to regulating galaxy formation, and track the growth
of baryonic galaxies within a full hierarchical structure formation
scenario.  These simulations have been fairly successful at modeling
post-reionization high-redshift galaxies \citep[e.g.][]{nig05,fin05},
so they provide a natural framework to extend studies back into the
reionization epoch.  Such an approach is complementary to models of first
star formation \citep{abe00,bro04} at $z\ga 15$, which aim to predict the
properties of primeval galaxies by understanding early stellar assembly
from primordial gas.

A significant difficulty with using models or simulations tuned to
match lower redshifts is that the physics of galaxy formation may be
quite different during the epoch of reionization.  Compared to today,
the universe during the reionization epoch had higher densities, shorter
cooling times, lower metallicities, and a flatter power spectrum on mildly
nonlinear scales, all of which could influence the formation and early
evolution of galaxies.  Furthermore, the process of reionization itself
may have large effects on the galaxy population \citep{bark01}, perhaps
suppressing star formation through photoionization heating \citep{bark99}
or stripping \citep{sig05}, or perhaps enhancing it through shock-induced
star formation or through early black holes emitting X-rays that stimulate
molecular hydrogen production \citep{hai05}.  It is therefore possible
that our recipes and models for galaxy formation tuned to observations at
the present epoch may have little relevance during the reionization epoch.
Indeed, this is almost certainly the case for primeval galaxies, i.e.
the very first galaxies forming nearly metal-free stars.

On the other hand, if we want to understand the properties of galaxies
that will be observable in the near future with next-generation
instruments such as {\it JWST} and 20-30m class telescopes, it is
questionable whether these first galaxies forming their very first
stars are the relevant objects to investigate.  In this paper we argue
that galaxies that will be detected in the foreseeable future have
already formed many generations of stars, as they are among the largest
and/or most vigorously star forming galaxies at these epochs.  They have
circular velocities well above that expected to be strongly affected by
photoionization, metallicities well above the threshold for a ``normal"
stellar initial mass function, and halo virial temperatures well above the
threshold where atomic lines become the dominant coolant.  Furthermore,
these systems are highly biased, which means that they are expected to
be relatively unaffected by the global reionization process occuring in
the universe at large, because they likely reionized their local volume
at an earlier epoch.  Hence we argue that observable galaxies at $z\la
9$ are probably more similar to lower-redshift systems than they are to
early primeval systems, and that simulation methods developed to model
post-reionization galaxy formation may provide a reasonably realistic
description for these systems that are responsible for completing the
process of reionization.

In this paper we study the detailed properties of the galaxy population
at $z\sim 6-9$ using cosmological hydrodynamic simulations.  We include
physical processes such as kinetic feedback with accompanying metal
injection and metal-line cooling, in addition to the usual radiative and
star formation processes required for galaxy formation, while ignoring
the highly complicating effects of radiative transfer.  Our primary goal
here is to use such models to make predictions for reionization-epoch
galaxies that can be tested against present and upcoming observations.
We find that many galaxies with stellar masses exceeding $10^8 M_\odot$
are already in place at $z=9$, and by $z=7$ there are plenty with
$M_*>10^9 M_\odot$.  These galaxies have all been enriched at significant
levels, to at least 3\% solar, and more typically 10\% solar.  They are
highly biased and highly clustered, and simple estimates show that they
will reionize their infall regions by $z\sim 9$ even if the rest of
the universe remains mostly neutral.  They would be detectable in large
numbers with near-infrared surveys only modestly deeper than accessible
today, and depending on line transfer effects could also be detectable
in \lya\ emission, perhaps even more abundantly than with broad-band
surveys.  We compare with observed $z\sim 6-7$ systems, and demonstrate
that our models nicely reproduce available observations.  In short,
the aim of this paper is to set the stage for detailed comparisons of
$z\ga 6$ objects with predictions from hierarchical models of structure
formation including gas physics, and show that initial comparisons can
already provide valuable insights into the nature of galaxies in the
reionization epoch.

Our paper is organized as follows:  In \S\ref{sec:sims} we present our
simulation methodology, details of our various runs, and algorithms for
determining galaxy properties.  In \S\ref{sec:phys} we present physical
properties of $z=6-9$ systems, including their masses, star formation
rates, metallicities, and halo properties.  In \S\ref{sec:clustering}
we discuss galaxy clustering and its evolution, along with implications
for local reionization.  In \S\ref{sec:observable} we present predictions
for broad-band and \lya\ emission properties of high-redshift galaxies,
including comparisons to available observations and prospects for
detection with future instruments.  We summarize our results in
\S\ref{sec:summary}.

\section{Simulations}
\label{sec:sims}

\begin{deluxetable}{lccccc}
\footnotesize
%\tablecaption{Simulation parameters.}
\tablewidth{0pt}
\tablehead{
\colhead{Name\tablenotemark{a}} &
\colhead{$L$\tablenotemark{b}} &
\colhead{$\epsilon$\tablenotemark{c}} &
\colhead{$m_{\rm SPH}$\tablenotemark{d}} &
\colhead{$m_{\rm dark}$\tablenotemark{d}} &
\colhead{$M_{\rm *,min}$\tablenotemark{d,e}}
}
\startdata
W8n256 & $8$ & $0.625$ & $0.484$ & $3.15$ & $15.5$ \\
W16n256 & $16$ & $1.25$ & $3.87$ & $25.2$ & $124$ \\
W32n256 & $32$ & $2.5$ & $31.0$ & $201$ & $991$ \\
\enddata
\tablenotetext{a}{Additionally, a suffix ``nw", ``cw", ``mzw" is added to denote no, constant, or momentum winds, respectively.}
\tablenotetext{b}{Box length of cubic volume, in comoving $\hmpc$.}
\tablenotetext{c}{Equivalent Plummer gravitational softening length, in comoving $\hkpc$.}
\tablenotetext{d}{All masses quoted in units of $10^6M_\odot$.}
\tablenotetext{e}{Minimum resolved galaxy stellar mass.}
\label{table:sims}
\end{deluxetable}

We employ the parallel cosmological galaxy formation code \gad\
\citep{spr05} in this study.  This code uses an entropy-conservative
formulation of smoothed particle hydrodynamics (SPH) along with a
tree-particle-mesh algorithm for handling gravity. It includes the
effects of radiative cooling assuming ionization equilibrium with
primordial chemistry, which we have extended to include metal-line
cooling based on the collisional ionization equilibrium tables of
\citet{sut93}.  The metal cooling function is interpolated to each
gas particle's metallicity as tracked self-consistently by \gad, and
added to the primordial cooling rate.  Gas particles that are eligible
for star formation are continually enriched based on an instantaneous
recycling approximation with a yield parameter of 0.02.  The thermal
energy of each gas particle is evolved on its cooling timescale (if it
is shorter than the Courant or dynamical timescale) assuming isochoric
conditions.  For full details see \citet{opp06}.  Stars are formed
using a recipe that reproduces the \citet{ken98} relation, employing a
subresolution multi-phase model that tracks condensation and evaporation
in the interstellar medium following \citet{mck77}; see \citet{spr03a}
for details.  Stars inherit the metallicity of the parent gas particle
when spawned, and from then on cannot be further enriched.

\gad\ has an observationally-motivated prescription for driving
superwinds out of star forming galaxies, which is tuned to reproduce
the global stellar mass density at the present epoch \citep{spr03}.
In \citet{opp06} we describe modifications to this scheme designed to
better match observations of IGM enrichment.  The free parameters are
the wind velocity of material ejected and the mass loading factor, which
is the ratio of the rate of matter expelled to the star formation rate.
Here, we will concentrate on three superwind schemes.  The first is one
where superwinds are turned off (though thermal feedback from supernova
is included).  This ``no winds" (nw) model is known to overproduce the
amount of stars significantly at low redshifts \citep{spr03}, so its
purpose here is to gauge the effects of the other (more observationally
consistent) superwind models.  The second is a ``constant wind" (cw) model
where all the particles entering into winds are expelled at $484\kms$
out of star forming regions and a constant mass loading factor of 2 is
assumed, which is the scheme used in the runs of \citet{spr03}.  Finally,
we introduce a new ``momentum wind" (mzw) model where the wind velocity
is proportional to the local circular velocity ($v_w=3*v_{\rm circ}$,
where $v_{\rm circ}$ is computed from the local gravitational potential),
and the mass loading factor is inversely proportional to the velocity
dispersion ($\eta=300\kms/\sigma$), as expected in momentum-driven wind
scenarios \citep{mur05} and consistent with observations by \citet{mar05}.
The mometum wind model also gives a velocity boost in low-metallicity
systems, based on the arguments that more UV photons are produced per
unit stellar mass at lower metallicities \citep[specifically, we employ
eqn.~1 of][]{sch03}, and that it is these UV photons that are driving
the wind \citep{mur05}.  Note that these supernova-driven winds must be
included ``by hand" mainly because we lack the numerical resolution to
self-consistently follow the physical processes required to drive bulk
flows out of star forming regions.

These superwind models are primarily intended to illustrate the
sensitivity of our results to our somewhat ad hoc feedback prescriptions,
though admittedly one could envision a wider range of feedback models.
We note that in \citet{opp06} we find the momentum wind model to be in
better agreement with a range of observations including intergalactic
\ion{C}{iv} absorption from $z\sim 5\rightarrow 1.5$, and our comparisons
in this paper also mildly favor the momentum wind model.  We further note
that observed typical outflow velocities for $z\sim 2-3$ Lyman break
galaxies of hundreds of km/s~\citep{pet01,sha03} and their inferred
mass loading factors of $\approx 4$ \citep{erb06} are both consistent
with our momentum wind model parameterization.  For these reasons we
generally focus on the momentum wind model when the conclusions from
the different wind models are broadly similar.

\gad\ implements heating due to a photoionizing background, which
we take from the latest CUBA model \citep{haa01} that assumes a 10\%
escape fraction of ionizing radiation from galaxies in addition to the
contribution from quasars.  For this model, reionization occurs
at $z\approx 9$.  Spatial uniformity is assumed, which is obviously
incorrect in detail during reionization, but may nevertheless be a
passable local approximation if the galaxies of interest reionize their
surroundings earlier and if their masses are significantly above that
which would be suppressed by photoionization.  For galaxies resolved by
our simulations, we will argue that both those statements are typically
true.

We run nine simulations:  Three different volumes as listed in
Table~\ref{table:sims}, each with our three different superwind schemes.
These three volumes can be used to judge numerical resolution effects
at overlapping mass scales.  We note that \gad\ has been extensively
tested for resolution convergence with regards to its star formation
and feedback algorithm~\citep{spr03a,nig05,fin05}, so we do not conduct
a wide suite of resolution tests.  Nevertheless we are reassured by
the good numerical convergence seen among our runs here, as we will
point out in upcoming sections.  The choice of box size has also
been carefully considered, because overly small volumes can lead to 
galaxy formation occuring too late owing to missing power on large scales
\citep{bark04}, while too large volumes would be unable to resolve the
early objects of interest.  Our box sizes are therefore chosen in a
range spanning between these pitfalls.

All runs use cosmological parameters consistent with recent joint
analyses of first-year WMAP, SDSS, Lyman alpha forest, and Type
Ia supernovae data \citep{teg04,sel04}:  We assume $\Omega=0.3$,
$\Lambda=0.7$, $H_0=70\kmsmpc$, $\sigma_8=0.9$, and $\Omega_b=0.04$.
More recent third-year WMAP data indicates a lower amplitude of the
matter power spectrum \citep[$\sigma_8\approx 0.75$;][]{spe06}, which
would result in fewer galaxies, but also a lower $\Omega$ and higher
$\Omega_b$ which would result in more galaxies at these epochs.  However,
there is still some debate on $\sigma_8$, as \citet{sel06} determines a
significantly higher value ($\sigma_8=0.85$) by incorporating Ly-$\alpha$
forest constraints.  We expect that our results would therefore be
qualitatively similar with the correct cosmology, and indeed there are
significantly larger uncertainties associated with modeling details such
as our superwind scheme.

Each of our runs has $256^3$ dark matter and $256^3$ gas particles,
evolved from the linear regime down to $z=6$.  The smallest volume is
a cube of $8\hmpc$ on a side, with an equivalent Plummer gravitational
softening length of $0.625\hkpc$ (comoving; i.e. better than 100~pc
physical at $z>6$), and our largest volume has a $32\hmpc$ box length.
The initial conditions are chosen to be random subvolumes of the universe,
generated using an \citet{eis99} power spectrum when the universe is still
well within the linear regime (typically $z\ga 150$).  Since the nonlinear
scale at $z=6$ is less than our smallest box length, each volume should
contain a representative galaxy population, though of course the rarest
objects will be absent.  Given the canonical value of 60 particles for
robust identification of dark matter halos \citep{wei99}, our smallest
volume is resolving halos with mass $\ga 2\times 10^8 M_\odot$, with
an associated virial temperature $\ga 10^4$~K.  As we will show in
\S\ref{sec:halos}, halos containing resolved galaxies (defined below)
are substantially larger, typically above $3\times 10^9 M_\odot$.
This means that primordial atomic gas should be the dominant coolant
in all our resolved systems.  Note that our simulations do not include
molecular hydrogen cooling.

\begin{figure}
\setlength{\epsfxsize}{0.65\textwidth}
\centerline{\epsfbox{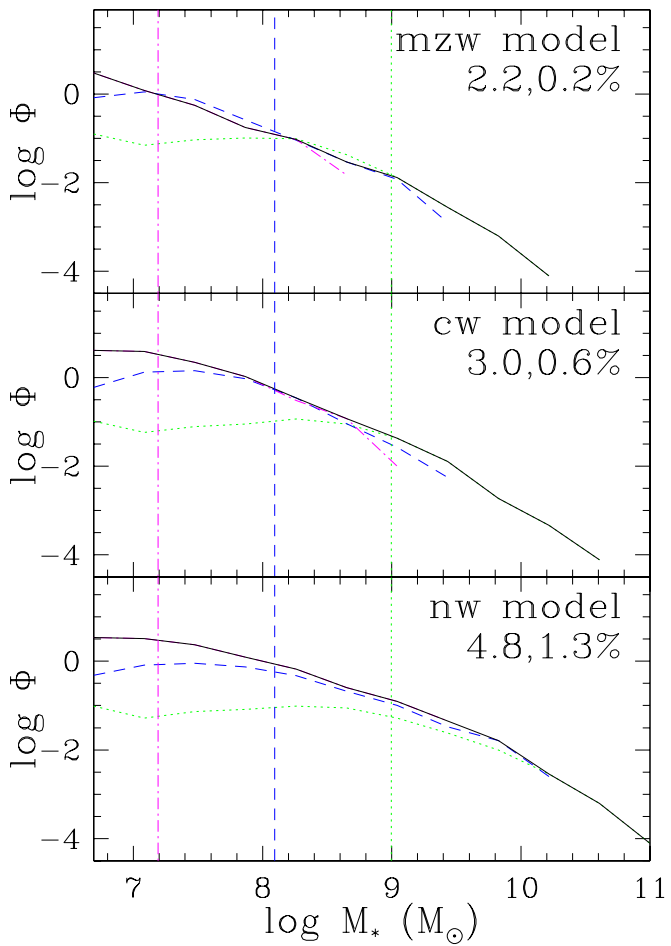}}
\vskip -0.5in
\caption{Stellar mass functions of galaxies at $z=6$ 
in the momentum wind (mwz) model (top panel), constant wind (cw) model
(middle panel), and no wind (nw) model (bottom panel).
Each panel shows the aggregate mass
function (solid line) as assembled from the individual mass functions
in the $8\hmpc$ box (dot-dashed line), $16\hmpc$ box (dashed line)
and $32\hmpc$ box (dotted line).  Vertical lines of the same line type
indicate the 64 star particle mass threshold for each volume.
The numbers in the upper right represent the percentage of 
baryons in galaxies and in stars, respectively.  The mass functions
join smoothly onto each other, showing good resolution convergence in the 
overlap region, and lending confidence to the procedure of stitching
together the mass functions.
}
\label{fig:mfcomp}
\end{figure}

We identify galaxies using Spline Kernel Interpolative DENMAX, and dark
matter halos using the spherical overdensity algorithm \citep[see][for
full descriptions]{ker05}.  We only consider galaxies with stellar
masses exceeding 64 star particles, which in \citet{fin05} we determined
represents a converged sample both in terms of stellar mass and star
formation history.  The minimum resolved stellar mass of galaxies in each
volume is listed in Table~\ref{table:sims}.  In order to fully utilize
the dynamic range in our three simulation volumes, we present stellar
mass functions derived by splicing together the results in our various
volumes, in the following manner:  For masses where multiple volumes
satisfy the 64 star particle criterion, we take the maximum value among
our simulated mass functions.  In practice, at any mass this is usually
the mass function of the largest volume (or close to it), and hence we
effectively present mass functions from the largest volume that resolves
a given galaxy mass.

The resulting $z=6$ stellar mass functions are shown in
Figure~\ref{fig:mfcomp}, for the momentum wind (top panel), constant wind
(middle) and no wind (bottom) models.  The dotted, dashed, and dot-dashed
curves represent the mass functions from the 32, 16, and $8\hmpc$ volumes.
Overlaid is the aggregate mass function, shown as the solid line, which
as expected tracks the $32\hmpc$ mass function at high masses and the
$8\hmpc$ mass function at small masses.  The vertical lines represent
the 64 star particle mass limit for the three volumes, which is where
the turndown begins, at least for the feedback models; for the no wind
model the turndown begins at slightly larger masses.  The caption in the
upper right shows the mass fraction of baryons in cold gas and stars at
$z=6$ in these three models, showing that the momentum wind model produces
significantly less cold gas and stars than the constant wind model by this
epoch, which in turn produces significantly less than the no-wind case.
The key point from Figure~\ref{fig:mfcomp} is that mass functions of
the different resolution simulations agree quite well in the overlapping
mass range, resulting in a nearly seamless mass function extending from
$10^{7.2}M_\odot$ to beyond $10^{10}M_\odot$.  Hence we do not expect
that numerical resolution limitations will affect our results, so long as
we concentrate on galaxies exceeding our 64 star particle mass criterion.

To obtain photometric properties of our simulated galaxies, we use
\citet{bru03} population synthesis models to convert the simulated star
formation history of each galaxy into broad band colours in an assortment
of filters, including $J$, $K_s$, and the {\it Spitzer} IRAC $[3.6\mu]$,
$[4.5\mu]$, $[8\mu]$ and MIPS $[24\mu]$ bands (we use AB magnitudes
throughout).  We describe this further in \S\ref{sec:spectra}.
We account for dust reddening using a prescription based on
the galaxy's metallicity and a locally-calibrated metallicity-extinction
relation.  We account for IGM absorption bluewards of rest-frame \lya\
using the \citet{mad95} prescription; at these redshifts this results
in virtually complete absorption, so we do not consider results from
bands at wavelengths shorter than redshifted \lya.  See \citet{fin05}
for a more complete description of these procedures.  We also predict
\lya\ emission line properties based on the instantaneous star formation
rates in each galaxy as output by \gad; we describe this in more detail
in \S\ref{sec:lya}.

We concentrate on studying galaxies between $z=6-9$ for several reasons.
First, our simulations have limited dynamic range, which means that
pushing to earlier epochs results in a statistically insignificant sample
of resolved galaxies that have formed by that epoch (particularly in
our larger volumes).  Second, our assumed ionizing background turns on
at $z\sim 9$, which means that prior to that time the suppression of
gas infall due to local photoionization, which may be relevant, is not
taken into account.  Finally, there is the practical consideration that
surveys of reionization galaxies are likely to concentrate on galaxies
in this epoch because they are both nearer and intrinsically brighter
than those at higher redshifts.

\section{Physical properties} \label{sec:phys}

\begin{figure}
\setlength{\epsfxsize}{0.7\textwidth}
\centerline{\epsfbox{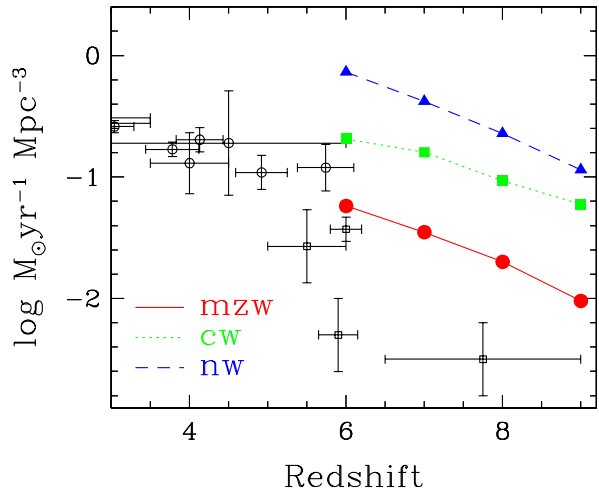}}
\vskip -3.5in
\caption{Global star formation rate density (Madau plot) for the
momentum wind (mzw; solid line), constant wind (cw; dotted line), and
no wind (nw; dashed line) models.  Open data points with error bars are
taken from a compilation by \citet{hop04}, which includes a star
formation rate-dependent extinction correction.  Open squares are
other data points at $z\geq 5.5$ from the literature: \citet[][$z=5.5$]{fon03},
\citet[][low point at $z\approx 6$]{bun04}, \citet[][high point at $z\approx 6$]{bou06}, and \citet[][$z\approx 7.8$]{bou04}.
The constant wind model
causes a suppression of $\approx\times 3$ versus no winds by $z=6$.
The momentum wind model provides further suppression by $\sim\times 3-4$
(depending on redshift) versus the constant wind case.  Extrapolations of
both constant and momentum wind models appear to be in agreement with
observations, while if no superwind feedback is implemented there is
significant overcooling.
}
\label{fig:madau}
\end{figure}

\begin{figure}
\setlength{\epsfxsize}{0.65\textwidth}
\centerline{\epsfbox{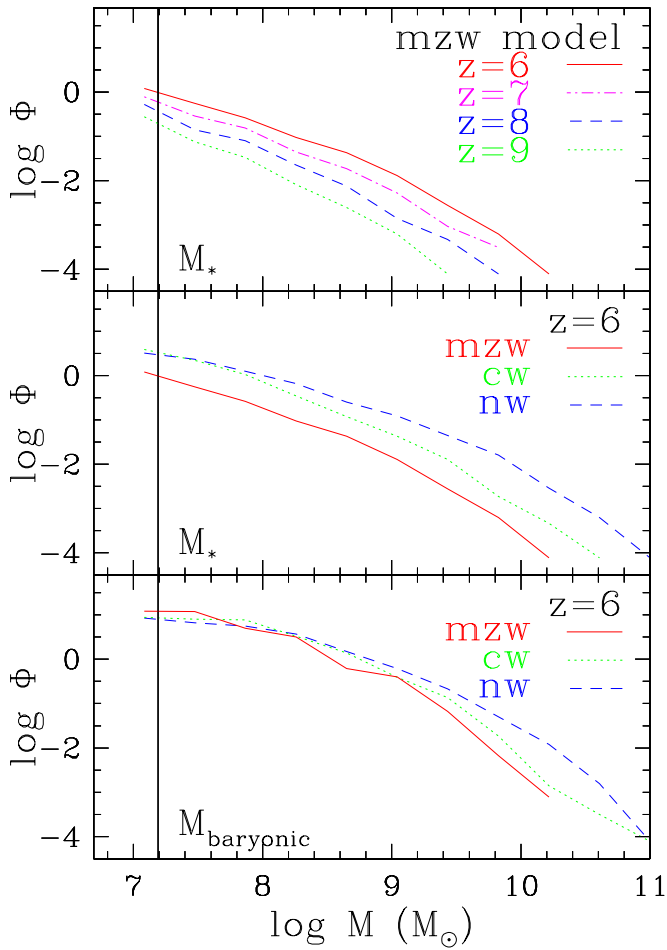}}
\vskip -0.5in
\caption{{\it Top:} Stellar mass functions of galaxies for $z=9, 8, 7, 6$
from the momentum wind model.
Units are number per cubic comoving $\hmpc$ per log stellar mass.
Vertical line is our galaxy stellar mass resolution limit of 64 
star particles.  As expected the stellar mass function increases with
time, and it additionally obtains a bend such that the faint-end 
slope is somewhat shallower with time.  
{\it Middle:} Stellar mass functions for the mzw (solid line), 
cw (dotted line), and nw (dashed line) models at $z=6$.  This shows
that the momentum wind model is most effective at suppressing star
formation across all galaxy masses.
{\it Bottom:} Same as middle panel, for total cold baryonic mass function
(i.e. star forming gas + stars).  The three
wind models have much more similar baryonic mass functions
as compared to their stellar mass functions.
}
\label{fig:massfcn}
\end{figure}

\subsection{Global Star Formation Rate Density}\label{sec:madau}

We begin by showing some basic physical properties of reionization-epoch
galaxies in our simulations.  The global star formation rate
density evolution, sometimes called the Madau plot \citep{mad96}, has
become a bellwether test for models of cosmological galaxy evolution.
Figure~\ref{fig:madau} shows the global star formation rate density for
the three wind models.  These values were computed from the resolved
galaxy population in the three volumes (8,16,$32\hmpc$) for each wind
model.  As found by \citet{spr03}, superwind feedback has a large effect
on suppressing star formation in the universe, even at these early epochs.
The constant wind case shows increasing suppression with time relative to
the no wind scenario, while the momentum wind model provides a relatively
redshift-independent global suppression of $\sim\times 10$.  Note that our
simulations include metal-line cooling, unlike those of \citet{spr03},
but at these early epochs primordial cooling is so efficient that the
additional cooling provided by metal lines does not significantly impact
the star formation efficiency \citep{her03,ker05}.

Figure~\ref{fig:madau} also shows data points from a compilation by
\citet{hop04} (only the $z>3$ data are shown).  We have supplemented
this with more recent measurements at $z\ga 5.5$ as noted in the
caption.  Comparing the simulations to the data, it is evident that
superwind feedback is required in order to bring the simulations into
broad agreement with observations and solve the overcooling problem
\citep{spr03}.  Based solely on these data it is not clear which of our
superwind models produces best agreement; indeed a simple extrapolation of
our two models seems to roughly bracket the observed range at $z\sim 3-6$.
The exceptions are the $z\approx 6$ point from \citet{bun04}, and the
$z\approx 8$ point from \citet{bou04}, which both lie significantly below
the predictions.  However, given that these points are based only on a handful
of bright objects detected at those redshifts, it is unclear what to make of
this discrepancy.  In general, observational selection effects could
significantly impact this comparison, as the galaxies identified
in our simulations may not be equivalent to those used to determine
the data points.  We leave more detailed comparisons for future work.
The main point here is that superwind feedback is a necessary ingredient
for suppressing early galaxy formation to obtain broad agreement with
observed cosmic star formation rate density evolution.

\subsection{Mass functions}\label{sec:massfcn}

Hierarchical structure formation models generally predict a rapid
early buildup of stellar mass in galaxies.  Stellar mass functions at
$z=9\rightarrow 6$ are shown in the top panel of Figure~\ref{fig:massfcn},
for the momentum wind model.  As expected, this function grows with time,
such that there are about twenty times more galaxies with $M\approx
10^9 M_\odot$ at $z=6$ as compared to $z=9$.  The shape is initially
close to a power law, and obtains a more pronounced bend with time; the
faint-end slope of a Schecter function fit goes from $-2.17$ at $z=9$
to $-1.85$ at $z=6$.  A significant number of galaxies with stellar
masses exceeding $10^8 M_\odot$, and a few exceeding $10^9 M_\odot$,
are already in place by $z=9$.  By $z=6$, galaxies exceeding $10^{10}
M_\odot$ have a space density of about $0.0002$ per cubic $\hmpc$.
This is comparable to the space density of the SDSS Luminous Red Galaxy
sample \citep{zeh05}, suggesting that these objects at $z=6$ are the
progenitors of large ellipticals at low redshift.

The middle and bottom panels show a comparison of stellar mass functions
and total galaxy baryonic mass functions for the various wind models.
The superwind feedback prescription has a significant effect on the
stellar mass function, producing nearly a factor of three reduction
in the mass function for the momentum wind model as compared with the
constant wind model.  The suppression of star formation in feedback
models relative to the no wind case is actually larger at larger masses,
counterintuitive to the notion that feedback preferentially affects
small galaxies \citep{dek86}; this reflects the hierarchichal assembly of
large galaxies from smaller ones.  There is much less of a difference among
models for the total baryonic mass function, because the constant wind
model has a relatively low mass loading factor that does not expel much
of the gas, while the momentum wind model has lower wind velocities that
allow fairly quick re-accretion.  The differences seen here can be regarded
as an estimate on the modeling uncertainty from our feedback algorithms.

\subsection{Stellar Birth Rates}\label{sec:sfr}

\begin{figure}
\setlength{\epsfxsize}{0.7\textwidth}
\centerline{\epsfbox{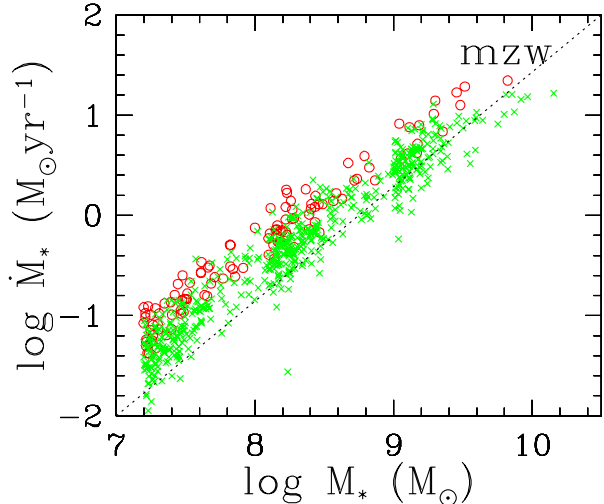}}
\vskip -3.5in
\caption{ Star formation rate as a function of stellar mass in the momentum 
wind model at $z=8$ (open circles) and $z=6$ (crosses).  The star
formation rate and stellar mass are generally proportional; the dotted line
shows a linear relationship for comparison.
}
\label{fig:sfrmstar}
\end{figure}

Figure~\ref{fig:sfrmstar} shows the star formation rates as a function
of stellar mass at $z=6$ and $z=8$ in the momentum wind model.
The three clumps of points correspond to galaxies from our three
volumes, which smoothly connect with each other.  Star formation rates
are essentially proportional to the stellar mass at both redshifts,
with the correlation being marginally tighter at $z=8$, implying
a fairly constant stellar birth rate across all masses.  This tight
relationship between stellar mass and star formation rate is also seen
at lower redshifts \citep{fin05}, and arises because star formation in
high-redshift galaxies is dominated by gas infall on dynamical timescales
\citep{ker05} rather than more stochastic merger-driven bursts as in some
semi-analytic models \citep[e.g.][]{kol99}.  Given that these simulations
cannot resolve the detailed internal dynamics of mergers that lead to
dramatic short-lived starbursts, it is possible that the scatter in this
relationship is underestimated.  Nevertheless the basic trend that the
most rapid star formers are generally the largest galaxies remains a
fundamental prediction of hydrodynamic simulations of galaxy formation.

Galaxies at a given stellar mass are actually forming stars approximately
twice as fast at $z=8$ than at $z=6$.  Note that the {\it total} star
formation rate in resolved galaxies increases by a factor of three from
$z=8$ to $z=6$ since there are many more resolved galaxies at $z=6$.
These statements are also generally true for the constant and no wind models
(not shown).  The largest galaxies at both redshifts are forming stars in
excess of $10\; M_\odot/$yr, which is vigorous even by today's standards.
This indicates that the epoch of galaxy growth is already well underway
by $z\sim 8$.

The tight positive correlation between stellar mass and star formation
rate is a basic prediction of hydrodynamic simulations at high redshifts
that has yet to be verified.  Observations by \citet{pap05} suggest that
at redshift $z\sim 2$, such a relation is broadly in place, although the
scatter is much larger than observed in simulations.  Such a trend is
certainly not observed today, as massive galaxies presently are forming
stars very slowly if at all.  It remains to be seen if simulations can
quantitatively reproduce this so-called galaxy downsizing, though the
qualitative trend is a natural outcome in hierarchical galaxy formation
models \citep[see][for more discussion]{dav05}.

The importance of mergers in driving galaxy properties at high redshifts
remains hotly debated.  In addition to the issue of stochasticity in star
formation due to mergers as discussed above, there is also the issue of
feedback due to merger-related processes, including the impact of central
black hole growth and associated feedback \citep{dim05}.  AGN feedback
has the potential to drive out large quantities of gas from galactic
potentials \citep{spr05b}, an effect we are are not modeling here.
It is plausible that at these high redshifts, where the quasar population
is small \citep{fan04}, the impact of such processes will be minimal.
However, this bears future testing, and is noteworthy as another caveat
in modeling the evolution of reionization-epoch galaxies.

\begin{figure}
\setlength{\epsfxsize}{0.7\textwidth}
\centerline{\epsfbox{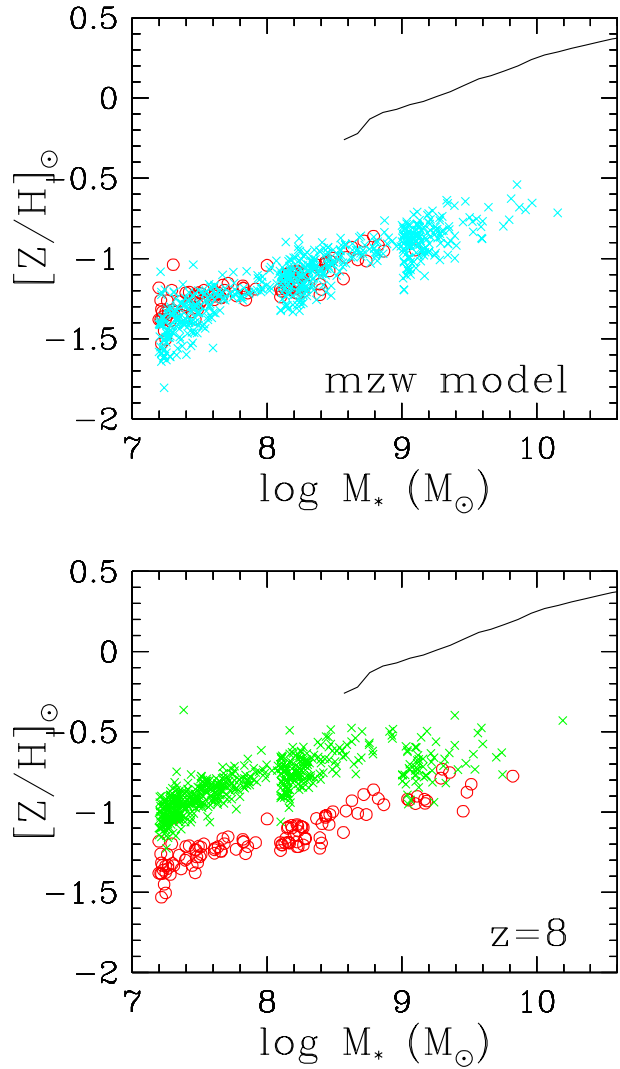}}
\vskip -0.5in
\caption{{\it Top:} 
Gas-phase metallicity versus stellar mass at $z=8$ (circles)
and $z=6$ (crosses) in the momentum wind model.  The three
``clumps" of points correspond to the three different simulation
volumes.  There is little evolution in the mass-metallicity relation 
at these early redshifts.  Line shows the median mass-metallicity
relation at $z\sim 0.1$ from SDSS \citep{tre04}.
{\it Bottom:} Gas-phase metallicity of galaxies at $z=8$
versus stellar mass, for the momentum wind (circles) and constant
wind (crosses) models.  The factor of three difference is
due to a similar difference in the efficiency of converting gas
into stars, as can be seen in the bottom panel of Figure~\ref{fig:massfcn}.
}
\label{fig:metevol}
\end{figure}

\begin{figure}
\setlength{\epsfxsize}{0.7\textwidth}
\centerline{\epsfbox{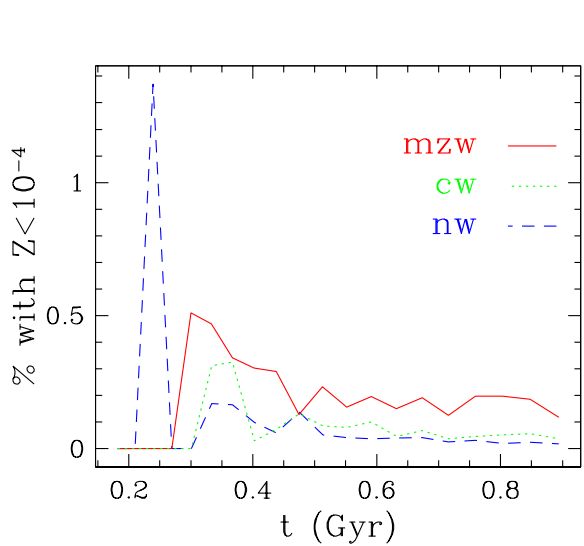}}
\vskip -3.5in
\caption{Percentage of stellar mass formed in our $8\hmpc$ simulations
with metallicity below $10^{-4}$ solar, as a function of time.
The fraction is always quite low, well below 1\% at all epochs in
the wind simulations, and the fraction generally falls with time.
This suggests that Population III star formation will be rare at $z\sim
6-9$, and even rarer at later epochs.
}
\label{fig:starmet}
\end{figure}

\begin{figure}
\setlength{\epsfxsize}{0.6\textwidth}
\centerline{\epsfbox{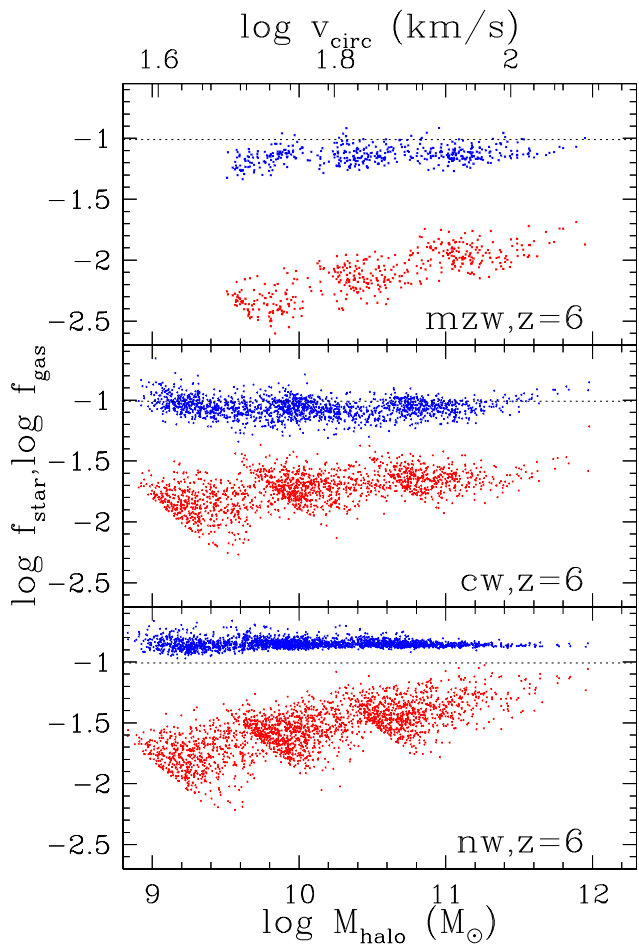}}
\vskip -0.5in
\caption{
Fraction of halo mass in total baryons (upper band) and stars
(lower band) as a function of halo mass at $z=6$ in the momentum wind
(top panel), constant wind (middle panel) and no wind (bottom panel)
models.  With no outflows, halos contain more than the global fraction
of baryons, indicated by the horizontal dashed line.  Outflows reduce
the baryon fraction relative to the no wind case by around 30\% and
20\% in the momentum and constant wind models, respectively.  The
stellar fractions a stronger trend with halo mass, reflecting
primarily the later collapse time of smaller halos.  The momentum wind case
shows a significant suppression of star formation relative to the other
two models, owing to the high mass loading factors in these small early
systems.
}
\label{fig:halofrac}
\end{figure}

\subsection{Metallicities}\label{sec:metals}

Metal pollution has a large effect on the reionization epoch, both in
terms of the cooling curve that strongly affects the Jeans mass in star
forming regions \citep{bro04}, as well as enabling the formation of dust
that can affect the detectability of systems, particularly in the \lya\
emission line.  In the top panel of Figure~\ref{fig:metevol} we show
the mass-averaged gas-phase metallicity of galaxies in the momentum
wind model at $z=8$ and $z=6$.  The mass-metallicity relation shows
little redshift evolution during this epoch; as galaxies form more
stars, they commensurately increase their metallicity so as to preserve
a linear relation.  For comparison, the SDSS median mass-metallicity
relation is shown as the solid line \citep{tre04}; interestingly, the
slope is similar, but the amplitude is almost 1~dex lower than observed
at $z\sim 0$.  This means that although little evolution is detected
from $z=8\rightarrow 6$, the metallicity at a given stellar mass must
increase by $\sim\times 5-10$ by redshift zero.  Given the protracted
period of time from $z=6\rightarrow 0$, this does not seem unreasonable,
but still bears future examination.  More recently, \citet{erb06} has
measured the mass-metallicity relationship at $z\approx 2$, and found
it to lie only 0.3~dex below the local relation; this is consistent with
the relatively slow evolution predicted in these simulations.

The bottom panel of Figure~\ref{fig:metevol} shows the gas metallicities
in the two wind models at $z=8$.  The feedback mechanism has a
significant effect on the metallicity, with the mometum wind model having
typically half a dex lower metallicity than the constant wind case.
This is directly a reflection of the $\sim\times 3$ lower amount of
stellar mass in galaxies of a similar total baryonic mass, as shown
in Figure~\ref{fig:massfcn}.  While we broadly favor the momentum wind
model, we note that the factor of 3 variation in the mass-metallicity
relation amplitude is probably indicative of current feedback modeling
uncertainties.  We have not shown the stellar metallicities, but they are
similar to the gas metallicities, being lower by at most a factor of two.

The main point of Figure~\ref{fig:metevol} is that in all cases, the
metallicities of resolved galaxies exceed about one-thirtieth solar.
While this would be considered metal-poor by today's standards, it
still far exceeds the metallicity threshold of $\sim 10^{-4}$ that is
purported to enact the transition to a normal Population II stellar
population from a ``metal-free" Population III one \citep[see][and
references therein]{bro04}.  Hence it is expected that these high
redshift galaxies should already have little if any contribution from
Population III stars, assuming that the metal mixing timescale in the
ISM is shorter than a few dynamical times.  Furthermore, these galaxies
may also contain a significant amount of dust if the timescale for dust
formation is as short, as has been inferred from high-redshift quasar
observations \citep{mao04}.  

While the mean metallicity in galaxies is large compared to the canonical
value required to suppress Population III star formation, it is still possible
that some stars continue to form with very low metallicities at late
times.  For instance, \citet{jim06} interpreted recent data from $z\sim
3-4$ galaxies as implying significant ($10-30\%$) ongoing Population III
star formation.  If verified, this would suggest that $z>6$ galaxies should
be forming even greater fractions of low-metallicity stars.

In Figure~\ref{fig:starmet} we calculate the percentage of stellar mass
formed in our $8\hmpc$ simulations having $Z<10^{-4}Z_\odot$ as a function of
time.  We choose our smallest volume to maximize the likelihood of early,
metal-free stars; the values from larger volumes were found to be lower.
The percentages are nonzero but very low, typically less than 0.5\%
at all epochs, with the only exception being the no wind case where
unfettered infall leads to an initial burst of near-metal free star
formation (still only $\sim 1.5\%$).  The stochastic nature of our star
formation algorithm, and the fact that we are not forming individual
stars but rather star ``clumps" that are $\approx 2\times 10^5 M_\odot$,
means that the values shown should be regarded as crude estimates.
However, it is still clear that our simulations produce very little
Population III star formation at $z=6-9$, likely dimishing further to
lower redshifts, and are therefore inconsistent with the \citet{jim06}
hypothesis.  Future observations will be required to determine if this
represents a failing of our models.

\subsection{Halo Baryon Fractions}\label{sec:halos}

During the reionization epoch, galaxies are likely embedded in a
inhomogenously percolating radiation field.  Due to the computational
complexity of radiative transfer, our simulations do not track this,
and instead assume a uniform photoionizing background turned on at
$z\approx 9$.  Clearly this approximation is incorrect in detail; however,
we can assess whether the resulting galaxy population is likely to be
significantly affected by the external ionizing radiation by examining
some basic properties of baryons in halos.

Figure~\ref{fig:halofrac} (top panel) shows the mass fraction of
total baryons (upper band of points) and stars (lower band) in halos
that host galaxies above our 64-particle stellar mass resolution
limit, for our momentum wind runs.  There are three clumps of points
corresponding to halos in our three volumes; the agreement between
the runs at overlapping masses indicates good resolution convergence.
Halos generally contain slightly less than the global baryon mass fraction
of $\Omega_b/\Omega_m=0.13333$, indicated as the horizontal dotted line.
There is a weak trend for smaller halos to contain a smaller baryon
fraction, as the momentum wind model increases the amount of material
ejected in these smaller systems.  However, even the smallest halos that
host a resolved galaxy still show a modest 30\% reduction in the baryon
fraction relative to the global mean.  The stellar fraction shows a more
clear trend with halo mass, indicating that our momentum wind model has
more of an effect at suppressing star formation than it does at ejecting
baryons from halos, as we saw in Figure~\ref{fig:massfcn}.

The middle and bottom panels show the same plot for the constant and
no wind models, respectively.  Unlike the momentum wind case, these
model shows no trends with halo mass in their baryon mass fractions,
and a reduced trend in their stellar mass fractions.  So, for instance,
in the lowest mass halos, there are more than three times as many stars
in the constant wind model than in the momentum wind case.  Again,
this can be taken as some measure of modeling uncertainty.

In all cases, halos containing resolved galaxies have masses $\ga 5\times
10^9 M_\odot$ and circular velocities $\ga 50$~km/s.  The circular
velocity scale is shown on the top axis, calculated as $v_{\rm 200}$
for a \citet{nav97} profile; the peak circular velocity would be higher
by 20--30\% assuming they follow an NFW-like rotation curve.  This means
that most of these galaxies are unlikely to be suppressed significantly by
photoionization \citep{tho96}, regardless of whether the photoionization
field is accurately modeled or not.  This is also evident in the baryon
fractions; had gas infall been suppressed due to photoionization heating,
the baryon fraction would be significantly reduced in smaller systems.
We have examined smaller volume ($4\hmpc$ box size) simulations and we
do in fact see significant suppression of the baryon fraction in halos
with $M_{\rm halo}\la 10^{8.5} M_\odot$, which is smaller than any halo
with a resolved galaxy in these runs.  It should be pointed out that
these galaxies and halos grew from smaller systems that may have been
more strongly affected by a local radiation field (if it was present) in
the past, but for particularly the more massive galaxies it appears they
will have spent most of their growth phase in a regime that is broadly
unaffected by whether or not a local photoionizing radiation field exists.

\section{Clustering and Inhomogenous Reionization} \label{sec:clustering}

Galaxy clustering is an important ingredient in the reionization process.
Early galaxies form within highly biased peaks in the matter distribution,
that are highly clustered.  This may enhance the inhomogeneity of the
reionization process, and may even result in a characteristic bubble size
for cosmological HII regions in the early universe \citep{wyi04,fur06}.
In this section we discuss clustering of our high-redshift galaxies,
and the implications for local reionization.

\subsection{Evolution of Clustering} 

\begin{figure}
\setlength{\epsfxsize}{0.55\textwidth}
\centerline{\epsfbox{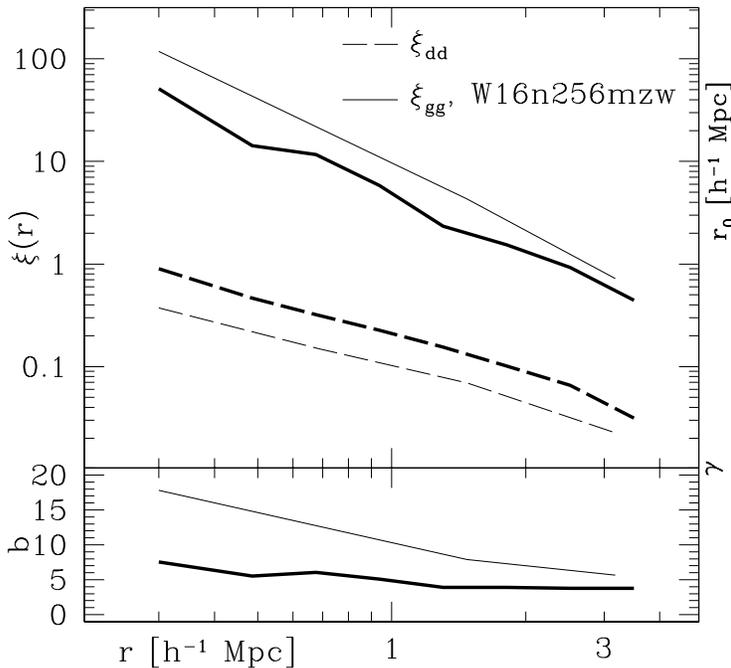}}
\caption{{\it Top:} Two-point correlation function for dark matter (dashed)
and galaxies with stellar mass $M_* > 10^{8\mbox{--}9} M_\odot$ (solid) 
from the $16\hmpc$ momentum wind model.  
Thin and thick curves indicate $z=9$ and 6, respectively.
{\it Bottom:} Bias $b\equiv (\xi_{gg}/\xi_{dd})^{1/2}$ as a function of
scale in the same model.  Galaxies reside in highly biased
regions, and the bias is a strong function of redshift and scale. }
\label{fig:xi}
\end{figure}

\begin{figure}
\setlength{\epsfxsize}{0.55\textwidth}
\centerline{\epsfbox{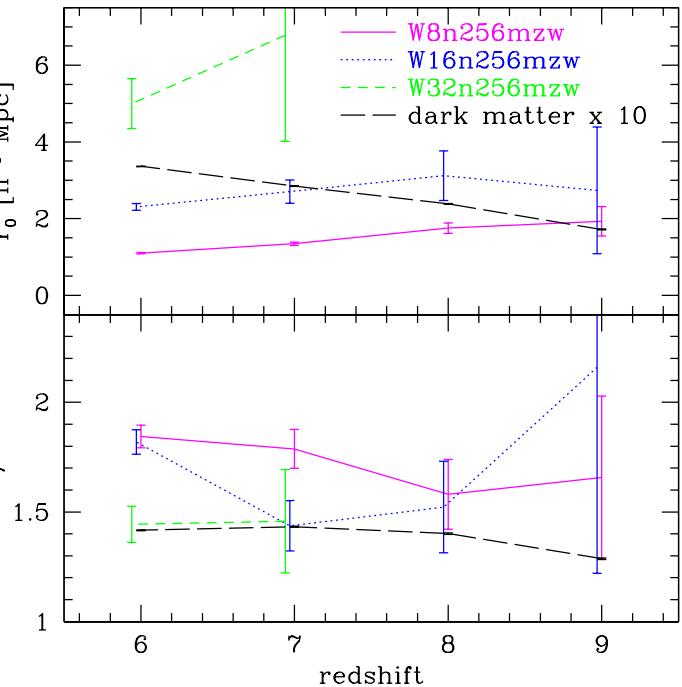}}
\caption{{\it Top:} Comoving correlation length versus redshift for dark
matter (long-dashed) and resolved galaxy samples from our three momentum
wind volumes.  The stellar mass ranges for galaxies in the volumes
in increasing order of size are $M_* \approx 10^{7.2\mbox{--}8} M_\odot$
$M_* \approx 10^{8.1\mbox{--}9} M_\odot$, and $M_* \approx 10^{9\mbox{--}10}
M_\odot$.  The dark matter correlation length has been multiplied by
a factor of 10 for visibility.  The error bars are Poisson, and do not
include cosmic variance.  They have been offset slightly for clarity.
Despite an increase of a factor of two in the dark matter correlation
length, the galaxy correlation length is roughly constant from
$z=9\rightarrow 6$.
{\it Bottom:} Slope of the correlation function versus redshift for the same
samples.  The slopes are broadly similar to that seen locally.}
\label{fig:r0}
\end{figure}

Figure~\ref{fig:xi} (top panel) shows the two-point correlation
function from the $16\hmpc$ momentum wind model computed for dark
matter ($\xi_{dd}$, dashed) and galaxies with stellar mass $M_*\approx
10^{8.1\mbox{--}9} M_\odot$ ($\xi_{gg}$, solid).  Thin and thick lines
indicate $z=9$ and 6, respectively.  The redshift evolution of $\xi_{gg}$
for mass-selected galaxy samples is determined by a competition
between the formation of new galaxies in less biased areas and the
gravitational growth of matter clustering.  As noted at lower redshifts
by~\citet[e.g.][]{wei04}, this can have the result that $\xi_{gg}$ is
nearly constant with redshift even though the comoving correlation length
of the underlying dark matter increases significantly.  This effect can
also be seen in the bottom panel of Figure~\ref{fig:xi}, which gives
the redshift evolution of the bias $b \equiv \sqrt{\xi_{gg}/\xi_{dd}}$.
As expected, the bias of the mass-selected sample drops dramatically
between $z=9$ and $z=6$, with the largest evolution occurring at the
smallest scales.

We have fitted power laws of the form $\xi(r) = (r/r_0)^{-\gamma}$ to
the correlation functions for the resolved galaxies and dark matter
in each of our simulations.  Figure~\ref{fig:r0} gives the redshift
evolution of $r_0$ for the 8, 16, and $32\hmpc$ volumes of the momentum
wind runs, with approximate mass ranges as indicated in the caption.
The dark matter correlation length is also plotted, multiplied by 10 to
increase visibility.  We have not attempted to measure the correlation
function of the resolved galaxies in the $32\hmpc$ box at $z=8-9$ because
this simulation contains too few resolved galaxies at these redshifts.
As anticipated by peak-bias formalism analyses, early galaxies are highly
clustered, with a strongly scale-dependent and mass-dependent bias.
Galaxy correlation lengths evolve only weakly with redshift, despite the
fact that dark matter correlation length doubles from $z=9\rightarrow
6$.  This suggests that the cumulative effects of photoionization from
collections of galaxies may significantly accelerate the reionization
process in local regions.  In the next section we estimate the strength
of this effect.

\subsection{Overlapping Galaxy Str\"omgren Spheres} \label{sec:overlapping}

\begin{figure}
\setlength{\epsfxsize}{0.5\textwidth}
\centerline{\epsfbox{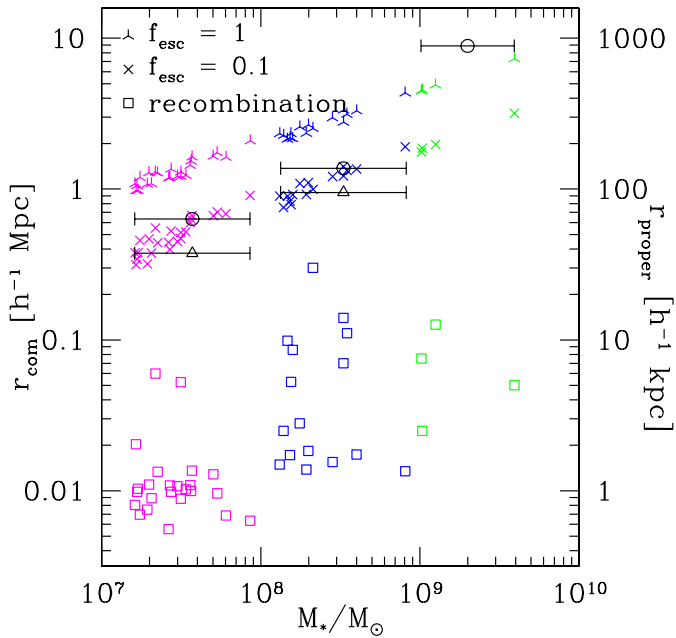}}
\caption{Radius of Str\"omgren sphere versus stellar mass for galaxies at
$z=9$.  The y-axes are in comoving (left) and proper (right) units.
Three-pointed and four-pointed stars denote the size of the ionized
IGM regions assuming escape fractions $f_{\mbox{esc}}=0.1$ and 1,
respectively, while the open squares show the radius of a region that
can recombine in a Hubble time.  Open circles and open triangles give
different estimates for the overlap radius as described in the text.
Galaxies less massive than $10^9 M_\odot$ have probably already achieved
overlap with a similar galaxy and ionized their infall regions by $z=9$
assuming $f_{\mbox{esc}}\geq 0.1$, while more massive galaxies have
probably achieved overlap with less massive galaxies.
}
\label{fig:ionize}
\end{figure}

Before reionization completes, the moment at which the ionized region
around a galaxy overlaps with a neighboring galaxy's ionized region
depends on the rate at which the galaxy leaks ionizing photons into its
surrounding IGM as well as the distance to the next galaxy.  In this
section we estimate the size of the ionized regions and compare this to
typical distances between galaxies, estimated using a neighbor search
as well as from the correlation function, in order to determine whether
galaxies' Str\"omgren spheres overlap by $z=9$.  Our calculation is
similar in spirit to that of \citet{bark02} and \citet{fur04}, with the
main difference being that we will use the gas distribution taken directly
from our simulations rather than one based on peak-bias formalism.

In order to estimate how much of the surrounding IGM each galaxy ionizes,
we determine how many ionizing photons the galaxies had emitted up to
$z=9$ by combining the simulated star formation history of each galaxy
with a fit to the number of ionizing photons emitted in the stellar
models of~\citet[equation 1]{sch03}.  These models assume a Salpeter initial
mass function (IMF)
between 1--100 $M_\odot$ and account for the effects of metallicities
down to near metal-free populations.  Typically, with this prescription
we find that 1~$M_\odot$ of stars formed by $z=9$ ionizes $\approx10^4
M_\odot$ of gas.  This is more than twice the ionized mass that would
be expected for solar metallicities ($\sim 4000 M_\odot$), in accord
with \citet{sch03}, which in turn in somewhat more than that expected
for a standard Salpeter IMF \citep{mad99}.  From the number of ionizing
photons we determine the mass of gas that would have been reionized
assuming a hydrogen mass fraction X=0.75, escape fractions $f_{esc}=$
1 and 0.1, and no recombination (we estimate recombination radii later).
Finally, we determine the radius of the sphere around each galaxy in the
simulation that encloses this mass of gas, including the galaxy itself.
This ionized sphere radius represents a crude estimate of the extent to
which a galaxy can ionize its surrounding IGM.

Figure~\ref{fig:ionize} shows, as a function of stellar mass, the
size of the ionized sphere around each resolved galaxy in the three
momentum wind runs.  Triangles and squares correspond to $f_{esc}=1$,
$f_{esc}=0.1$, respectively; as we discuss below, the latter is probably
more relevant, while the former is included to show the sensitivity
to this uncertain parameter.  These points closely follow a slope
$r\propto M_*^{1/3}$.  Since the ionizing flux is closely tied to the
stellar mass, which in turn is reasonably closely tied to the halo mass
(cf. Figure~\ref{fig:halofrac}) this indicates that the universe is
fairly homogeneous at these radii.  In words, this plot indicates that
galaxies with stellar mass around $10^{8.5} M_\odot$ are capable of
ionizing spheres around $\approx 1 h^{-1}$ Mpc (comoving) in size,
assuming an escape fraction $f_{esc} \geq 0.1$; regions ionized by
lower-mass galaxies scale in size with the galaxy's stellar mass roughly
as $r\propto M_*^{1/3}$.

Do the ionized regions overlap by $z=9$? A brute-force approach to
this question is to determine distance to the nearest neighbor for
each galaxy in each simulation, take the median, and divide by 2.
This distance, which we call the overlap radius $R$, is indicated by the
large open circles in Figure~\ref{fig:ionize}; the error bars indicate
the mass ranges that the samples cover.  A slightly more elegant
approach is to find the radius $2R$ of the sphere that is expected
to enclose one neighbor using the known mean number density $n_0$ and
galaxy autocorrelation function: $\int_0^{2R} n_0 (\xi(r) +1) dV = 1$.
Using this approach, we obtain overlap radii given by the open triangles,
which are somewhat smaller than that obtained using our brute-force
approach.  No estimate is made for the largest mass bin because there
are too few galaxies at $z=9$ in the $32\hmpc$ run to reliably estimate
a correlation function.

Comparing the size of the ionized regions to either overlap radius
for each sample indicates that galaxies with stellar masses
$10^{7\mbox{--}9} M_\odot$ have already achieved overlap with a similar
galaxy by $z=9$.  By contrast, the galaxies in the most massive sample
are too rare to have achieved overlap with a similar galaxy at $z=9$
despite the large ionized regions that they can create.  By $z=8$, $R$
drops to $3.5 h^{-1}$ Mpc for these galaxies so that they do overlap.
Nevertheless, since these massive galaxies fall in highly biased
regions, it is reasonable to expect that by $z=9$ they will already
have achieved overlap with nearby lower-mass galaxies \citep[as argued
by e.g.][]{fur04,wyi05}.  This means that their infall region, i.e. the
region of space where gas is falling onto that galaxy as opposed to any
other, has likely been ionized by then.

Of course, the actual physical situation is considerably more complicated
than the simple model we have presented here.  First of all, ionizing
photons must escape from the galaxy as well as the surrounding dense
IGM in which the ionized hydrogen would be able to recombine quickly.
An estimate of size of this region is shown as the open squares in
Figure~\ref{fig:ionize}.  This is computed as the radius of the sphere
which encloses the mass of gas (including the galaxy) which, had it
been ionized very early on, would have been able to recombine within a
Hubble time.  The recombination time becomes very small close to the
galaxy, so in order for the IGM to become ionized, some fraction of
the ionizing flux from each galaxy must be permitted to escape beyond
this radius.  A canonical value for this escape fraction is $\sim 10\%$
\citep[cf.][]{haa01}, motivating our above choice for this parameter.
Our analysis shows that once an ionizing photon does manage to escape into
the IGM, the recombination time becomes comparable to a Hubble time and
most of the ionized gas can be expected to remain ionized.  Recombinations
will also reduce the size of the Str\"omgren sphere somewhat, but for
these bubble sizes the effect should not be large \citep{fur05}.

Further complications that we do not consider in this model are the
non-sphericity of the matter distribution that results in a more complex
morphology for ionized regions, the fact that the universe is denser
at earlier times and therefore the recombination region may actually be
somewhat larger, and effects of light-travel time and photon redshifting.
A full cosmological radiative transfer hydrodynamics code is required
to model all these effects, which is beyond the scope of this paper.
We note that a number of radiative transfer calculations for the early
universe have already been done \citep[e.g.][]{gne00,raz02,cia03,sok04,ili05},
so we hope to include full modeling of the topology of reionization
in the future.  But for the purposes of this work, our quasi-analytic
modeling is likely to be qualitatively accurate.

In summary, our simulations combined with simplistic modeling suggest
that, by $z=9$, galaxies more massive than $10^7 M_\odot$ are likely to
have achieved ionized bubble overlap with other galaxies even if the
reionization process has not yet completed in less biased regions.
These galaxies should have therefore ionized their infall regions
prior to $z\sim 9$.  The overlapping Str\"omgren spheres should contain
within them a photoionizing background strength that is comparable to
that estimated by \citet{haa01}.  Hence our assumption of a uniform
photoionizing background, while not valid for the bulk of the volume,
may be reasonably accurate for the regions around the galaxies that we
are interested in.

\section{Photometric and Emission Line Properties of Galaxies} \label{sec:observable}

\begin{figure}
\setlength{\epsfxsize}{0.7\textwidth}
\centerline{\epsfbox{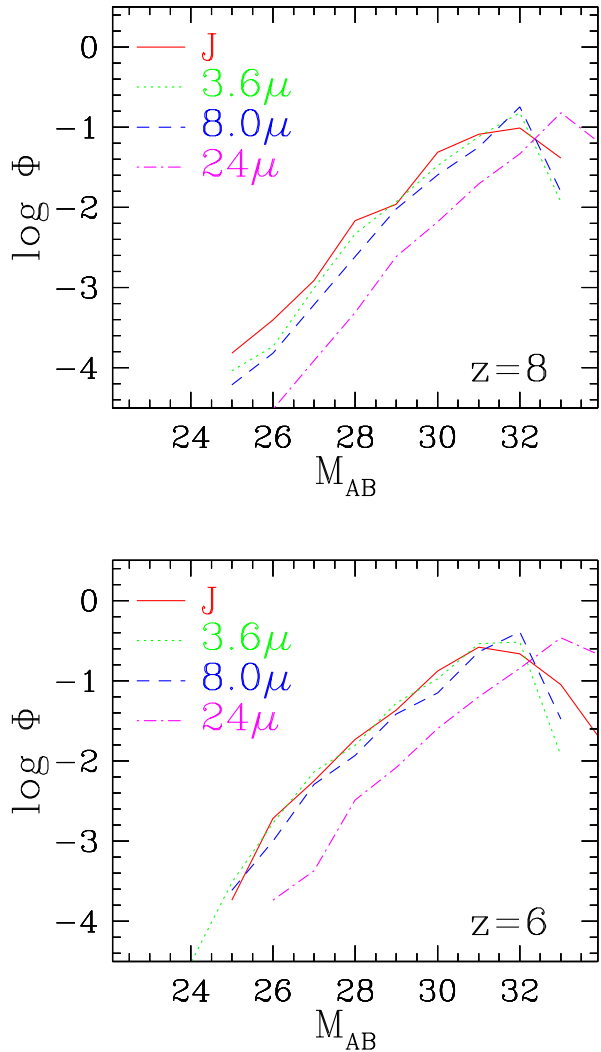}}
\vskip -0.5in
\caption{Luminosity functions of galaxies in the momentum
wind model for various observed IR bands at $z=8$ (top panel) and.  
$z=6$ (bottom panel).  The luminosity functions are fairly steep,
and show that galaxies are generally blue, becoming bluer at higher
redshifts.  
}
\label{fig:lumfcn}
\end{figure}

\begin{figure}
\setlength{\epsfxsize}{0.7\textwidth}
\centerline{\epsfbox{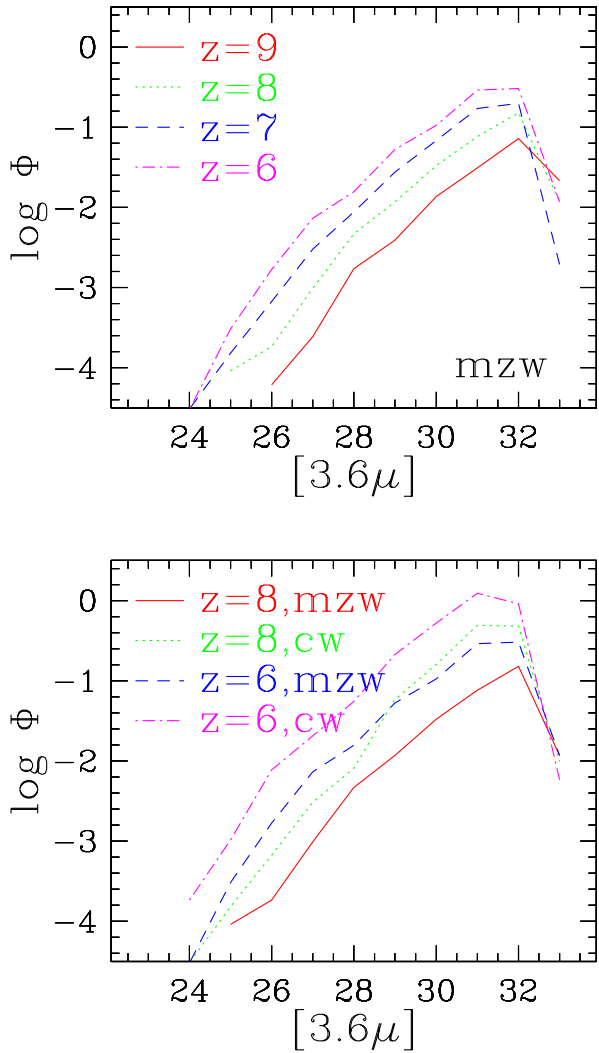}}
\vskip -0.5in
\caption{{\it Top:} Evolution from $z=9\rightarrow 6$ of the luminosity 
functions of galaxies in the momentum wind model in the {\it Spitzer} 
$[3.6\mu]$ band.  
{\it Bottom:} Comparison of $[3.6\mu]$ luminosity functions in the 
constant wind and momentum wind models, at $z=8$ and $z=6$.  The 
constant wind model has higher luminosities and a steeper faint-end slope.
Units of $\Phi$ are number per $(\hmpc)^3$ per magnitude.
}
\label{fig:lumfcnz}
\end{figure}

\begin{figure}
\setlength{\epsfxsize}{0.55\textwidth}
\centerline{\epsfbox{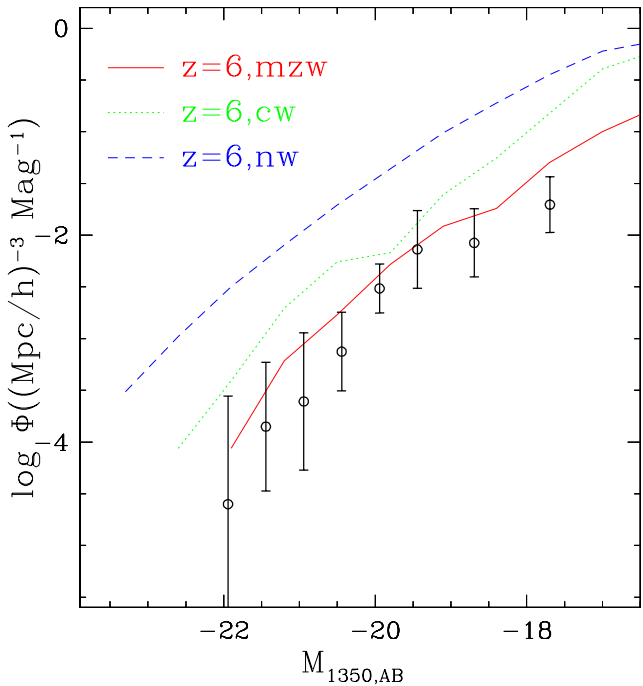}}
\vskip -1.1in
\caption{Comparison between observations of the rest-1350\AA\ luminosity function
of $z\sim 6$ $i$-dropout galaxies by \citet{bou06} versus our three wind models.
The momentum wind model predictions are in good agreement with data,
the constant wind model predicts about three times too many galaxies,
and including no superwind feedback results in more than an order of 
magnitude too many galaxies.
}
\label{fig:lfabs}
\end{figure}

High-redshift galaxies have mainly been discovered and observed using
two techniques: Broad-band colour selection and \lya\ emission line
searches.  Broad band selection, usually utilizing the Lyman continuum
edge or even the \lya\ edge at high redshifts, offers the advantage
of requiring only photometric surveys.  However, galaxies are faint in
these broad bands, particularly reionization-epoch objects which require
infrared observations that face high sky backgrounds from the ground.
Also, the precise redshift cannot be determined, hampering three-dimensional
clustering and environmental studies.  \lya\ selection has
the advantage that sources may be very bright in young, dust-poor systems
as expected in the early universe, they can be found in narrow-band
filters that offer low sky background, and a fairly precise redshift is
immediately obtained.  However, \lya\ is a difficult line to interpret,
as it is easily scattered and extincted, so that even a smattering
of dust can result in highly attenuated flux.  A full understanding of
reionization epoch galaxies will likely require pursuing both approaches.
In this section we present the broad band and \lya\ emission properties
of $z=6-9$ galaxies from our simulations.

\subsection{Photometric Properties}\label{sec:photo}

We examine a series of bands redwards of redshifted \lya\ at $z>6$,
since bluewards bands are expected to be heavily attenuated by intervening
neutral hydrogen.  Figures~\ref{fig:lumfcn} and \ref{fig:lumfcnz}
show various luminosity functions (LFs) for the two wind models for
$J\rightarrow [24\mu]$ bands, corresponding roughly to rest-frame 1500\AA\
to $K$-band at $z\sim 6-9$.  These LFs have been obtained by stitching
together the individual LFs from our three volumes for galaxies above our
resolution limit, in a manner similar to that described for mass functions
in \S\ref{sec:sims}.  There is generally good agreement in the overlap
regions resulting in a smooth LF from 24th to 32nd magnitudes, although
the overlap regions are typically less than a magnitude.  The turnover
at faint magnitudes arises from our 64 star particle mass resolution limit
in our smallest volume.

Figure~\ref{fig:lumfcn} shows the $z=8$ and $z=6$ predicted LFs in four
infrared broad bands.  Galaxies are generally blue, with the rest-$K$
LF lower by about two magnitudes at both redshifts.  Hence while
galaxies possess an old stellar population, their light is dominated by
young stars.  Galaxies at $z=8$ are somewhat bluer than at $z=6$,
as at $z=8$ the rest far-UV LF (observed $J$) is larger than the rest
blue LF (observed $[3.6\mu]$), which is larger than the rest optical
(observed $[8\mu]$) one; by $z=6$ the LFs for all these bands are similar.
Galaxies as bright as 25th magnitude in near-IR bands exist at $z=8$, but
they are quite rare and presumably highly clustered, requiring large-area
surveys to find.  The steep LFs suggest that modest increases in depth
will yield substantially larger samples.

Figure~\ref{fig:lumfcnz} (top panel) shows the evolution of the Spitzer
$[3.6\mu]$ band LF from $z=9\rightarrow 6$ in the momentum wind model.
The LF shows a steady increase, by about three-quarters of a magnitude
per unit redshift, slowing a bit from $z=7\rightarrow 6$.  At $z=9$
the LF looks like a power law, but by $z=6$ it obtains a slight bend,
akin to the stellar mass functions in Figure~\ref{fig:massfcn}.  Similar
evolution is seen in other bands.

The bottom panel shows a comparison of the $z=8$ and $z=6$ LFs between
the momentum wind and the constant wind models.  The constant wind model,
by virtue of its greater amount of star formation, has higher luminosity
functions.  The shape is also slightly affected, with the constant wind
model showing a steeper faint-end slope.  This is a reflection of the
stellar fraction varying more with mass in the momentum wind model than
in the constant wind case, as shown in Figure~\ref{fig:halofrac}; the
high mass loading factor in small galaxies is suppressing the faint-end
slope in the momentum wind model.  This may prove advantageous if the
trend continues to lower redshifts, since simulated luminosity functions
from constant wind simulations at $z\sim 4$ appear to have too steep a
faint end compared to observations \citep{nig05,fin05}.

Observations of $z\sim 6$ galaxies are already rapidly accumulating.
\citet{bou06} found over 500 systems in {\it Hubble} NICMOS and ACS
parallel fields that satisfy $i$-dropout Lyman break selection criteria.
Of course, some are probably interlopers, but they estimate a rather
small level of contamination ($\la 8\%$).  Figure~\ref{fig:lfabs} shows
a comparison of our $z=6$ luminosity functions computed at rest-frame
1350\AA\ to their observations.  Overall, the agreement in both shape and
amplitude is quite impressive, particularly for the momentum wind model
(solid line).  Note that no parameters have been tuned specifically to
obtain this agreement.  The no wind model (dashed line) overpredicts
the number of galaxies at a all luminosities by well over an order
of magnitude, showing that feedback must suppress galaxy formation
at {\it all} masses in order to match observations.  \citet{bou04}
detected a handful of $z\sim 7-8$ systems down to $H<28$, and found that
the evolution from $z\sim 4$ is about a factor of five in number; we
predict (cf. Figure~\ref{fig:lumfcnz}) a factor of $\sim 3$ increase from
$z\approx 8\rightarrow 6$, so the observed evolution is not inconsistent
with our models, though this bears further examination down to $z=4$.

While this comparison favors the momentum wind model at face value, it is
worth recognizing that the number of high-redshift galaxies is sensitive
to the CDM power spectrum on small scales, which is still not precisely
constrained.  For instance, early galaxy formation can be significantly
suppressed in warm dark matter \citep{yos03a} or running spectral index
\citep{yos03b} models.  Since these $\Lambda$CDM variants remain viable at
present \citep[though increasingly disfavored;][]{sel04}, we do not claim
to constrain our feedback models based on $z\ga 6$ galaxy abundances.
Furthermore, the new WMAP3 data favors a slightly lower matter power
spectrum at these redshift at the relevant scales, which would lower
our predicted LFs.  Hence we leave a more quantitative study of this
for future work.  Here we simply note that our models broadly agree
with existing observations of $z\sim 6$ LFs, and with improvements in
cosmological parameter precision such data could play a key role in
constraining models of early galaxy formation.

\subsection{Sample Galaxy Spectra}\label{sec:spectra}

\begin{figure}
\setlength{\epsfxsize}{0.55\textwidth}
\centerline{\epsfbox{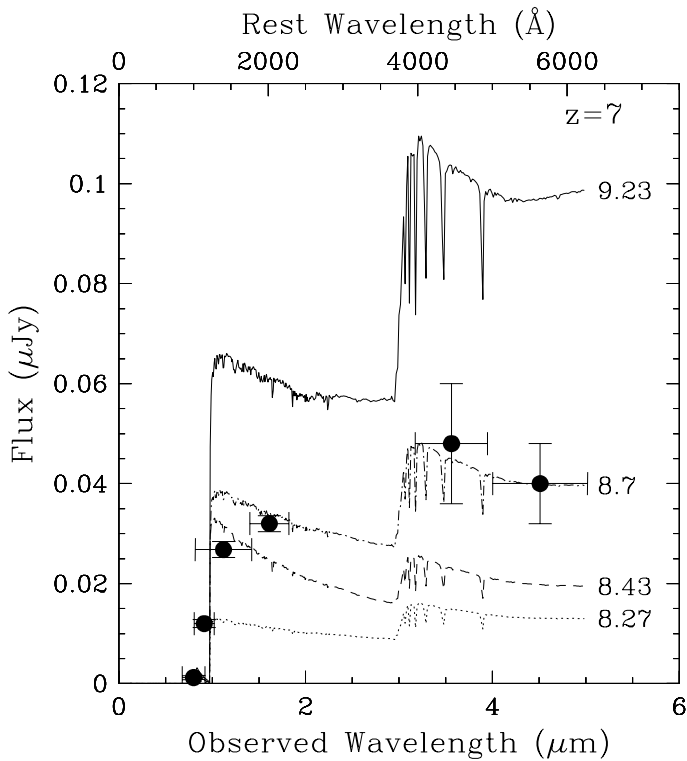}}
\vskip -1.1in
\caption{Sample galaxy spectra drawn from the $16\hmpc$ momentum wind
simulation.  These generally show a significant 4000\AA\ break and a
blue continuum, with some variation due to differences in star formation
histories.  The log of the stellar mass of each galaxy is shown on the
right.  The top one is the most massive galaxy in this run at this epoch,
and the rest are randomly selected within various mass ranges.
The data points are observations of a $z\approx 7$ galaxy by \citet{ega05}
corrected by a factor of 25 for lensing magnification.  The data are
well fit by one of our galaxies with $10^{8.7}M_\odot$.
}
\label{fig:spectra}
\end{figure}

Thanks to unprecedented sensitivities of {\it Spitzer} and other 
near-infrared instruments, it has now become possible to study the
stellar populations of galaxies at $z>6$ at some level of detail through
multi-band photometry spanning into the rest-optical.  This provides even
more stringent tests on our simulated galaxy population.  Here we examine
the rest-UV and optical spectra for a few of our simulated galaxies.
These spectra are obtained from \citet{bru03} population synthesis
models using the Padova 1994 models and a Chabrier IMF, where each
star particle in a galaxy is treated as a single stellar population
whose age is computed from the star particle's time of creation.
The single-burst stellar population spectra of a given galaxy are then
summed to obtain that galaxy's spectrum.  We interpolate the population
synthesis models to the stellar metallicity of each galaxy.  Note that
we are not employing a top-heavy IMF, since the metallicities of these
galaxies fall within the range where there are no compelling arguments
for deviations from locally-derived IMFs.

Figure~\ref{fig:spectra} shows a selected sample of spectra from
the $16\hmpc$ simulation at $z=7$.  As can be seen, most of these
systems possess a significant old stellar population as evidenced by
a significant 4000\AA\ break, and have a continuum that is typically
slightly blue through the rest-UV, consistent with the LFs shown in
Figure~\ref{fig:lumfcnz}.  There is some variation in the continuum
shape due to the details of individual star formation histories,
but overall there is no trend with mass, because the birthrates of
galaxies are similar at all masses (cf. Figure~\ref{fig:sfrmstar}).
Indeed, these galaxies look not terribly dissimilar to present-day star
forming galaxies, with the exception of their low luminosity (and of
course no emission lines since these are not present in the Bruzual \&
Charlot models).

Recently, \citet{ega05} observed a putative $z\sim 7$ object found though
a cluster lens search \citep{kne04} in {\it Spitzer} and {\it Hubble}
near-IR bands.  They showed that this system was best modeled as having
$\sim 5-11\times 10^8 M_\odot$ of stars with a substantial old stellar
population.  Their data, corrected downwards by a factor of 25 for lensing
magnification, are shown as the data points in Figure~\ref{fig:spectra}.
A galaxy randomly drawn from our $16\hmpc$ momentum wind simulation
having a stellar mass of $5\times 10^8M_\odot$ just so happens to be an
excellent match to these data.  This galaxy has a star formation
rate of $1.5\; M_\odot$/yr, a median stellar age of 130~Myr, and one-tenth
solar metallicity, which are all typical values for a galaxy of this size
in our simulations.  They are also consistent with values inferred from
a more detailed analysis of this object by \citet{sch05}.  This shows
that galaxies like the Egami et al. object are not only expected, but
are fairly common and typical systems at this epoch.

Another object, a $z=6.56$ lensed \lya\ emitter found by \citet{hu02},
was followed up with {\it Spitzer} by \citet{cha05}, who found that its
spectrum is similar in shape but intrinsically somewhat brighter than the
Egami et al. object.  Not surprisingly, it is possible to find simulated
galaxies with $M\sim 10^9 M_\odot$ that match most of the broad-band data
for this object as well.  However, this object does have a peculiarly
high IRAC $[4.5\mu]$ flux.  By fitting a model to the remaining data and
assuming the excess flux is due to unresolved H$\alpha$ line emission,
Chary et al. estimate a star formation rate of $\sim 140 M_\odot/$yr.
This star formation rate would be far above that expected for galaxies
of this mass (cf. Figure~\ref{fig:sfrmstar}), so if confirmed this would
present a challenge to our models.  However, this object was also analyzed
by \citet{sch05} prior to the {\it Spitzer} data becoming available,
and their estimated star formation rate is $11-41 M_\odot/$yr, much more
in line with our simulations.  Another object, the claimed detection
of a $6\times 10^{11}M_\odot$ system with a photometric redshift of
$z\approx 6.5$ by \citet{mob05}, would be difficult for our models,
or indeed any $\Lambda$CDM-based model, to reproduce, as the expected
space density of such systems is less than one per cubic gigaparsec.
However, other groups favor a substantially lower redshift (and lower
mass) for this object, so it remains to be seen if this galaxy will
indeed force revisions to our cosmology.

\subsection{\lya\ Emission}\label{sec:lya}

\begin{figure}
\setlength{\epsfxsize}{0.7\textwidth}
\centerline{\epsfbox{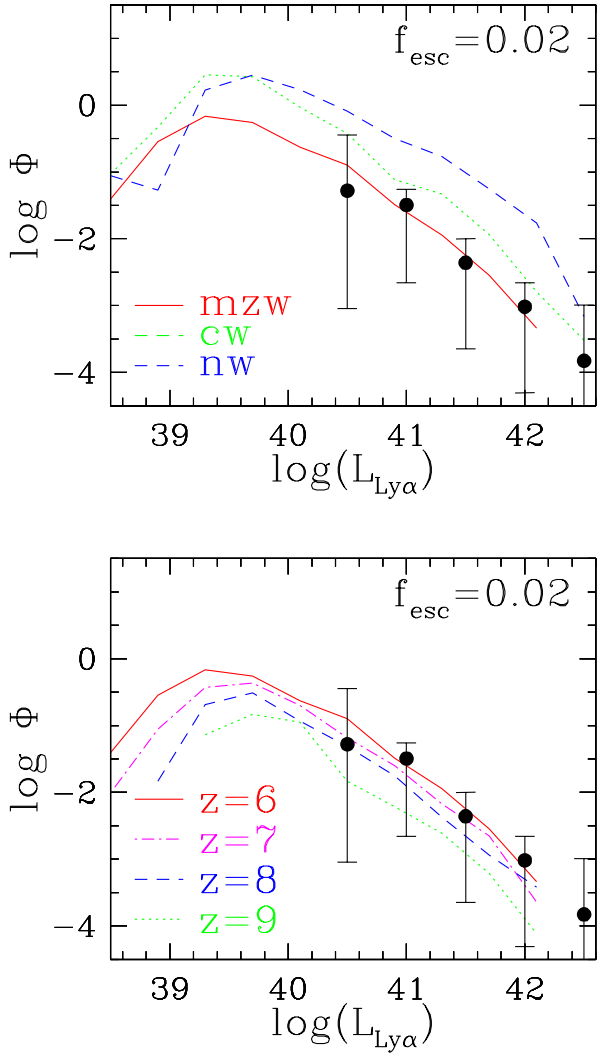}}
\vskip -0.5in
\caption{{\it Top:} The Lyman alpha emitter luminosity function at $z=6$
in the momentum wind (solid), constant wind (dotted), and no wind (dashed) models, assuming
a Ly$\alpha$ escape fraction of 2\%.  The units of $\Phi$ are number
per unit $(\hmpc)^3$ per log luminosity interval, and $L_{Ly\alpha}$ is
in erg/s.  Data from \citet{san04} are shown for comparison.  For $f_{\rm
esc}=0.02$ the mzw model matches reasonably well to
the data.  {\it Bottom:} The evolution of the \lya\ luminosity function
from $z=9\rightarrow 6$ in the momentum wind model, for an assumed escape
fraction of 2\%.  The evolution is about half an order of magnitude in
$L_{Ly\alpha}$ at a given number density over these redshifts.
}
\label{fig:lyalf}
\end{figure}

\lya\ emission galaxies are now being detected out to $z\approx 6.5$
using narrow-band surveys \citep{hu02,mal04,ste05}.  Higher redshifts move
into the infrared, where current space-based observatories do not have
sufficient collecting areas to detect these faint sources, necessitating
ground-based observations through a bright night sky.  Fortunately, there
are some relatively clean, albeit narrow, windows between water and OH
lines where the sky background is low \citep[see e.g.][]{bar04}.  A number
of narrow-band surveys are underway to take advantage of these windows to
search for \lya\ emitters at $z\sim 7-10$ \citep{bar04,hor04,wil05,sta05}.
In this section, we make predictions for the \lya\ emitter population
from our simulations for comparison with such observations.

As discussed in \citet{bar04}, there are large uncertainties regarding the
detectability of \lya\ emission from star formation. Scattering due to the
surrounding IGM including the damping wing may make \lya\ emission more
diffuse and difficult to detect.  As we showed in \S\ref{sec:metals},
significant metals are expected to be present, so dust extinction may
be considerable, unless the ISM remains surprisingly unmixed.  On the
other hand, if the initial mass function is extremely top-heavy, then
perhaps \lya\ emission can be boosted.

To make predictions for \lya\ luminosity functions from our simulations,
we convert the instantaneous star formation rate of each galaxy
into a \lya\ luminosity using a fit to the models listed in Table~4
of \citet{sch03}.  In particular, we use a solar metallicity \lya\
emission value of $2.44\times 10^{42}$~erg~s$^{-1}$$M_\odot^{-1}$~yr,
and then correct for individual galaxy metallicities using equation~1
of \citet{sch03}.  Of course, not all \lya\ photons emitted by young
stars will actually reach us; there is an escape fraction that is
highly uncertain.  Note that by ``escape fraction" here we mean the
fraction of \lya\ photons that reach our telescopes, not the fraction
that escapes from the star forming region or galaxy itself that is
canonically around 10\%.  The escape fraction depends on the amount of
scattering and absorption in the ISM and IGM along the line of sight,
which is difficult to predict directly.

Here we take the approach of constraining this escape fraction empirically
by comparing our $z=6$ predicted \lya\ luminosity function with data
from \citet{san04}.  They used Keck/LRIS to obtain spectra of objects in
high-magnification regions of foreground clusters, and found 11 likely (7
confirmed) \lya\ emitters with $z\approx 4.6-5.6$.  Lensing magnification
offers the best current technique for detecting faint sources, and indeed
these objects represent the faintest \lya\ emitters currently known.
The downside is that lensing results in a considerably smaller volume
being probed.

In Figure~\ref{fig:lyalf} we present simulated luminosity functions for
\lya\ emitters from the three wind models (top panel) and the evolution
from $z=9\rightarrow 6$ in the momentum wind model (bottom panel).
We choose an escape fraction of 2\% for \lya\ photons that would
be detectable by us,  which is selected to obtain agreement with the
\citet{san04} observations (using all 11 likely sources) in the momentum
wind case, shown as the data points.  For the constant wind case, an
escape fraction of 0.5\% is an equivalently good fit, and for the no wind
case the best-fit escape fraction is around 0.2\%.  Note that the escape
fraction tunes only the amplitude of the luminosity function, while the
slope remains an independent prediction of the models (assuming that the
escape fraction is not luminosity-dependent).  Therefore the excellent
agreement from $\log{L_{Ly\alpha}}=40.5\rightarrow 42$ is encouraging.
Though the Santos et al. data would be more appropriately compared
to $z\sim 5$ predictions, \citet{hu05} has demonstrated that there is
no discernible evolution in the \lya\ luminosity function from $z\sim
4\rightarrow 6.5$, so the comparison to our simulated $z=6$ sample should
be reasonably valid.

\citet{mal04} and \citet{hu05} also argue that the lack of evolution from
$z\approx 5.7$ to $z\approx 6.5$ would be surprising if the universe
was predominantly neutral at $z\approx 6.5$, since the damping wings
from IGM absorption would significantly reduce the flux and change
the line profiles.  Even if the bulk of the universe is neutral, it is
possible that the bright \lya\ sources seen at $z=6.5$ may have ionized
their local surroundings, facilitating the escape of \lya\ photons
\citep{hai02,san04}.  In either scenario, it appears \lya\ emission will
continue to escape from galaxies and be detectable into the reionization
epoch.  Our crude estimates of the local photoionization volume in
Figure~\ref{fig:ionize} supports the idea that even out to $z\sim 9$,
IGM damping wing encroachment may not significantly attenuate \lya\
emission, which agrees with conclusions reached by various semi-analytic
studies \citep{fur04,wyi05,mal05}.

If we take our estimated escape fraction at $z=6$ and assume that it does
not evolve with redshift, then there is relatively little evolution
predicted from $z=9\rightarrow 6$, as shown in the bottom panel of
Figure~\ref{fig:lyalf}.  This scenario is broadly consistent with Hu
et al.'s lack of observed evolution to lower redshifts.  At a given
number density, the evolution is around a factor of three in luminosity
between $z=9\rightarrow 6$, or equivalently a factor of five in number
density at a given luminosity.  This means that surveys designed to
probe $z\approx 9$ will still have plenty of candidate targets, they
must ``merely" go $3\times$ deeper than those at $z\approx 6$.

One instrument that is being currently developed to probe \lya\ emitters
in the reionization epoch is the Dark Ages $z$ \lya\ Explorer, or DAzLE
\citep{hor04}, set to begin operation in 2006.  Using an $R=1000$ survey
in J-band night sky windows on the Very Large Telescope, DAzLE will
survey 6.8'x6.8' patches down to $2\times 10^{-18}$~erg~s$^{-1}$cm$^{-2}$
in the redshift range $z=7.7-10$.  We can make predictions for the
$z=7.7$ atmospheric window, with a redshift range of $\Delta z=0.05$
over which the sky is relatively dark \citep[see][for a diagram of the
J-band night sky spectrum]{bar04}.  Here, their flux limit translates
to $10^{41.7}$~erg/s, and the window's volume for the DAzLE field is
$2500 \;(\hmpc)^3$.  Our predicted source density above this flux limit
is approximately $0.002\;(\hmpc)^{-3}$, so DAzLE should expect to see
around 5 sources per field within this night sky window.  The numbers are
similar for the $z=8.2$ window, which is a bit wider but further away,
and the $z=8.8$ window which has $\Delta z\approx 0.1$ but half as many
predicted sources than at $z=7.7$.  These numbers are not large, but with
a few adjacent fields it may be sufficient to do preliminary clustering
studies to possibly constrain the patchiness of reionization.  If no
objects are seen, this would suggest that the escape fraction is dropping
substantially to higher redshifts, whereas an abundance of objects would
signify a dramatic change in the IMF or dust content of $z\ga 6$ systems.
More accurate predictions for detailed comparisons will require the
development of \lya\ radiative transfer codes including dust extinction
\citep[e.g.][]{tas06}.

Another key design issue is the expected line widths of \lya\ emission.
Since halos at this epoch have circular velocities typically less than
$150\kms$ (cf. Figure~\ref{fig:halofrac}), the naive expectation is
that emission line widths will not be significantly larger than this.
It may be significantly smaller if the emission region (or last scattering
surface) is not sampling the full potential well, as seen e.g. for Lyman
break galaxies at $z\sim 3$ \citep{pet01}.  Conversely, it could be
larger if outflows stimulate \lya\ emission, or if radiative transfer
effects smear out the emission region beyond the galaxy's host halo,
though these scenarios seem relatively improbable.  Hence it is likely
(but not guaranteed) that an instrument with $R=1000$ such as DAzLE would
not resolve out \lya\ emission lines, which makes internal kinematic
studies difficult but is preferable for source detection.

In summary, the \lya\ emission properties of $z\ga 6$ galaxies are an
interesting way to constrain the topology and efficiency of reionization.
Our simulations suggest that \lya\ emitters will be observable out to
$z\sim 9$ with current technology, assuming that galaxies sufficiently
ionize their surroundings so that the escape fraction is similar to that
at $z\la 6$.  Detection of these objects appears feasible in the near
future, which should provide crucial insights into the reionization
epoch.

\subsection{Next Generation Instruments}

A key science goal of {\it JWST} is to image the objects responsible
for reionizing the universe.  The NIRCam team has a galaxy formation
program designed to image $z\ga 7$ galaxies \citep{rie03}.  The survey
will have 50ks exposures in six near-IR filters, and 100ks in their
4.4$\mu m$ filter.  The expected depths achieved will be roughly 1~nJy,
or 31.4~AB magnitudes, across these bands.  They then plan to do
follow-up emission line studies on selected high-$z$ objects.

Our simulated luminosity functions are coincidentally resolved down to
about 1~nJy, so we can make general predictions for the types of objects
that will be observed by such a survey.  For starters, the predicted
number density will be large: At $z\approx 8$, each 2.3'x2.3' field will
contain over 400 galaxies per unit redshift (cf. Figure~\ref{fig:lumfcn}),
which should be straightforwardly identifiable as B1-band dropouts
\citep[see][for filter set]{rie03}.  Note that while large, these numbers
are still smaller than previous estimates: The semi-analytic model
of \citet{bark00} predicts many thousands of galaxies down to 1~nJy.
The main difference is our suppression of star formation through
superwinds, which appears to be necessary in order to match observed
$z\sim 6$ luminosity functions (cf. Figure~\ref{fig:lfabs}).  We also
predict galaxies will be strongly clustered, but as each NIRCam field is
over $4\hmpc$ (comoving) on a side which is longer than the correlation
length (cf. Figure~\ref{fig:r0}), the field-to-field variations should
not be too large.  The physical properties of the observed galaxies are
likely to be what we have described so far: Mildly enriched systems living
in halos of masses $\ga 10^{9.5}M_\odot$, with the smallest stellar masses
observed being a few$\times 10^7 M_\odot$.  It is not expected that even
the faintest of these objects will push into the regime of predominantly
metal-free star formation with little or no dust, though such objects
may enter into the survey if their stars are exceptionally bright.
In short, these are not the primeval galaxies you're looking for.

Turning to \lya\ emitters, the expected faintness of these systems
requires larger collecting areas, and fortunately work is already underway
on the 22m Giant Magellan Telescope ({\tt http://www.gmto.org}),
among others.  Such a telescope with an $R=3000$ near-IR tunable
narrow-band capability would probe an order of magnitude fainter than
DAzLE~\citep{bar04} resulting in $\sim 20\times$ more sources, equivalent
in space density to $\sim 30$th magnitude objects.  Such spectral
resolution may enable kinematic studies as well.  Such capabilities
promise to move \lya\ emitter searches from a handful of discovery
objects into the realm of large statistical samples for detailed studies.

\section{Summary}
\label{sec:summary}

We have presented results for the physical and observable properties of
reionization epoch galaxies from cosmological hydrodynamic simulations.
Our simulations naturally produce an early epoch of star formation that
yields significant numbers of reionization-epoch galaxies, in broad
agreement with available observations.  We find that:

\begin{itemize}

\item Systems with $10^8 M_\odot$ in stars are common by $z\sim 9$, and
by $z=6$ systems with $10^{10} M_\odot$ have a similar space density to
luminous red galaxies today.  The mass functions have a steep faint end
initially, but become somewhat shallower in time, particularly in our
favored wind feedback model.

\item Star formation rates are well correlated with stellar masses,
such that the stellar birthrates show almost no trend with mass at these
epochs, modulo the impact of stochasticity in star formation caused
by merger-related processes.  At $z=8$, the stellar mass doubling time
for galaxies is around 0.2~Gyr, while at $z=6$ it is $\approx 0.4$~Gyr.
Star formation rates at $z\sim 9\rightarrow 6$ range from hundredths to
tens of solar masses per year.

\item The large stellar masses imply significant metal enrichment in these
systems.  While they are generally subsolar and can be comparable to the
lowest metallicity dwarfs today such as I~Zw~18, their metallicities are
still well above the putative threshold for ushering in a more normal
IMF as compared to a top-heavy IMF proposed for Population III stars.
We predict few if any near metal-free stars forming in these systems.

\item The halos containing these galaxies have masses well above
$10^9M_\odot$, meaning that their virial temperatures exceed the
primordial cooling limit of $10^4K$, and hence atomic lines are the
dominant coolant.  Their baryon content is somewhat below the global
average due to feedback, and different feedback recipes can produce
factors of three differences in the stellar content, depending on mass.
Their circular velocities are $\ga 50 \kms$, implying that these galaxies
are unlikely to be significantly affected by photoionization.

\item Galaxies at this epoch are highly biased and highly clustered, with
a strongly mass-dependent clustering strength and scale-dependent bias.
The rapid growth of matter clustering is offset by a rapidly dropping
bias, producing a relatively constant comoving correlation length.
Galaxies are sufficiently clustered that they are likely to lie in
regions that have been locally reionized at a much earlier epoch than
the time of final cosmological \ion{H}{II} region overlap.

\item The broad-band properties of these galaxies show that they are
generally fairly blue systems, getting bluer to higher redshifts.
Hence the optimal detection band will be the one that is just redwards
of \lya.  There is evolution of about $0.5-0.75$ magnitudes per unit
redshift in near-IR bands.  Predicted number densities vary by $\sim\times
3$ depending on the wind model, but the evolution is generally similar.

\item Predictions of \lya\ emitters are less robust because of large
uncertainties in \lya\ transfer out of these systems.  Assuming a
\lya\ emission detectability similar to that observed at $z\approx
5$, then surveys with expected sensitivities of future 20-30m class
telescopes will find an equivalent number density of $z=9$ objects as
near-infrared surveys down to $\sim 30$st magnitude AB.  Hence \lya\
narrow-band surveys can be a powerful tool for obtaining large samples
of reionization epoch objects.

\item Our simulations show good agreement with observations of $z\ga
6$ objects.  Individual $z\sim 7$ galaxy broad band spectra have been
observed \citep[e.g.][]{ega05} that are well-matched by simulated
galaxies, showing the requisite continuum slope and 4000\AA\ break.
Our $z\approx 6$ luminosity function agrees well with observations
by \citet{bou06}, particularly for our momentum wind model (modulo
uncertainties regarding the exact amplitude of the matter power spectrum).
The \lya\ luminosity function at $z\approx 6$ is in good agreement with
observations by \citet{san04}, for a reasonable \lya\ escape fraction
of 2\%.  We look forward to comparisons with future data to provide more
stringent constraints for our models.

\end{itemize}

The outlook for detecting large numbers of reionization epoch objects
is promising.  Current {\it Hubble} data can achieve depths of 28.5 in
near-IR bands \citep{bou06}, albeit in small fields.  In addition to {\it
Webb}, several future space-based telescopes have been proposed that will
do large-area near-IR surveys, including some versions of JDEM (Joint
Dark Energy Mission), as well as the proposed Galaxy Evolution Probe
(R. Thompson, PI).  Though these missions remain firmly in the distant
and uncertain future, the scientific richness of deep wide near-IR
surveys ensures that they are likely to be accomplished at some point.
On the \lya\ emitter side, a number of operational instruments should
begin detecting a handful of systems in the immediate future, but surveys
in near-IR night sky windows on future 20-30m class telescopes should
bring these studies to maturity.  The timescales for such facilities
are of the same order as the space-based infrared ones, so it will be
an interesting race to see which technique matures first.  The future
looks bright for lifting the veil of reionization and probing the final
frontier of observational cosmology.

\section*{Acknowledgements}

A catalog of resolved galaxies from the simulations presented
here, with photometry, spectra, and all physical properties, is
available upon request.  The simulations used here were run on the
Xeon Linux Supercluster at the National Center for Supercomputing
Applications.  This work was supported in part by {\it Hubble} theory
grant HST-AR-10647.01-A.  Partial support for this work, part of the
Spitzer Space Telescope Theoretical Research Program, was provided
by NASA through a contract issued by the Jet Propulsion Laboratory,
California Institute of Technology under a contract with NASA.  We thank
V. Springel and L. Hernquist for providing us with \gad.  We thank
E. Barton, C. Papovich, D. Eisenstein, N. Katz, D. Keres, M. Rieke,
J.-D. Smith, and D. Weinberg for helpful discussions.

\end{document}